\documentclass[times]{elsarticle}
\usepackage[margin=2.5cm]{geometry}

\usepackage{graphicx}
\usepackage{amssymb}

\usepackage{lineno}
\usepackage{graphicx}
\usepackage{amsmath}
\usepackage{amsfonts}
\usepackage{gensymb}
\usepackage{graphicx}
\usepackage{subfigure} 
\usepackage{hyperref}

\usepackage{floatrow}
\usepackage{amssymb}
\usepackage{latexsym}
\usepackage{url}
\usepackage{xcolor}

\DeclareMathOperator*{\argmin}{arg\,min}
\journal{Medical Image Analysis}

\begin{document}

\begin{frontmatter}


\title{Model-Informed Machine Learning for Multi-component $T_2$ Relaxometry}



\author[1,2]{Thomas {Yu}}

\author[1,4]{Erick Jorge Canales-Rodr\'{i}guez \corref{cor1}}
\cortext[cor1]{Corresponding author: erick.canalesrodriguez@epfl.ch ({Erick J Canales-Rodr\'{i}guez)}}

\author[5,1]{Marco Pizzolato}

\author[7,3,1]{Gian Franco Piredda}

\author[7,3,1]{Tom Hilbert}

\author[1,9]{Elda Fischi-Gomez}

\author[9,10]{Matthias Weigel}

\author[9,10]{Muhamed Barakovic}

\author[2,1,3]{Meritxell Bach Cuadra}

\author[9,10]{Cristina Granziera}

\author[7,3,1]{Tobias Kober}

\author[1,3]{Jean-Philippe Thiran}

\address[1]{Signal Processing Lab 5 (LTS5), \'{E}cole Polytechnique F\'{e}d\'{e}rale de Lausanne, Lausanne, Switzerland}

\address[2]{Medical Image Analysis Laboratory, Center for Biomedical Imaging (CIBM), University of Lausanne, Switzerland}

\address[3]{Department of Radiology, Lausanne University Hospital and University of Lausanne, Switzerland}

\address[4]{FIDMAG Germanes Hospitalàries Research Foundation, Centro de Investigación Biomédica en Red de Salud Mental (CIBERSAM), Barcelona, Spain}

\address[5]{Department of Applied Mathematics and Computer Science, Technical University of Denmark, Kongens Lyngby, Denmark}


\address[7]{Advanced Clinical Imaging Technology, Siemens Healthcare AG, Lausanne, Switzerland}


\address[9]{Translational Imaging in Neurology Basel, Department of Medicine and Biomedical Engineering, University Hospital Basel and University of Basel, Basel, Switzerland}

\address[10]{Neurologic Clinic and Policlinic, Departments of Medicine, University Hospital Basel and University of Basel, Basel, Switzerland}

\begin{abstract}
Recovering the $T_2$ distribution from multi-echo $T_2$ magnetic resonance (MR) signals is challenging but has high potential as it provides biomarkers characterizing the tissue micro-structure, such as the myelin water fraction (MWF). In this work, we propose to combine machine learning and aspects of parametric (fitting from the MRI signal using biophysical models) and non-parametric (model-free fitting of the $T_2$ distribution from the signal) approaches to $T_2$ relaxometry in brain tissue by using a multi-layer perceptron (MLP) for the distribution reconstruction. For training our network, we construct an extensive synthetic dataset derived from biophysical models in order to constrain the outputs with \textit{a priori} knowledge of \textit{in vivo} distributions. The proposed approach, called Model-Informed Machine Learning (MIML), takes as input the MR signal and directly outputs the associated $T_2$ distribution. We evaluate MIML in comparison to non-parametric and parametric approaches on synthetic data, an \textit{ex vivo} scan, and high-resolution scans of healthy subjects and a subject with Multiple Sclerosis. In synthetic data, MIML provides more accurate and noise-robust distributions. In real data, MWF maps derived from MIML exhibit the greatest conformity to anatomical scans, have the highest correlation to a histological map of myelin volume, and the best unambiguous lesion visualization and localization, with superior contrast between lesions and normal appearing tissue. In whole-brain analysis, MIML is 22 to 4980 times faster than non-parametric and parametric methods, respectively.
\end{abstract}


\end{frontmatter}


\section{Introduction}
The spin-spin relaxation rate $T_2$ is one of the basic tissue-specific, quantitative parameters which can be measured or used to give image contrast in MRI \citep{haacke1999magnetic}. However, while commonly presented as a single number per voxel, tissue heterogeneity and partial volume effects renders it more appropriate to consider distributions of $T_2$s per voxel rather than a single $T_2$ value \citep{menon1991application}. We distinguish single-component $T_2$ relaxometry, where each voxel is characterized with a single $T_2$, from multicomponent $T_2$ relaxometry, where each voxel is characterized with a $T_2$ distribution. In general, $T_2$ distributions are reconstructed from multi-echo $T_2$ MRI signals, which can be acquired, for example, through multi-echo spin echo sequences, where a 90\degree\ excitation pulse is followed by a train of 180\degree\  refocusing pulses. Given a sequence of $n$ pulses, the signal $\mathbf{s}$ is a vector of $n$ measurements at the corresponding echo times ($TE_i$). Let $p(T_2)$ and  $\alpha$ denote the distribution of $T_2$s in a voxel and the effective flip angle of the refocusing pulses, respectively. 
If $\alpha=180$ and the voxel is assumed to have a single $T_2$, then the decay of the signal is exponential, as is implied by the Bloch equations \citep{bloch1946nuclear}. In practice, inhomogeneities in the transmit field ($B_1+$) result in an effective refocusing pulse that can vary significantly from 180\degree\ and can be spatially heterogeneous \citep{prasloski2012applications}. This leads the resulting signal to deviate from the ideal exponential behavior, which can be modelled using the extended phase graph (EPG) formalism  \citep{hennig1988multiecho}. The EPG formalism considers as parameters $\alpha, TE$, $T_1$ and a single $T_2$. In this work, we use the common simplification of fixing $T_1=1000ms$, as the T1 relaxation time cannot be estimated using the acquisition sequences we examine in this work \citep{neumann2014simple}; hence, it is commonly fixed to its mean value in brain tissue. Then the normalized signal follows
\begin{align}
     \mathbf{s}(TE_i) = \int EPG(TE_i,T_1,T_2,\alpha) p(T_2) \mathrm{d}T_2.
\end{align}
One key application of multi-component $T_2$ relaxometry is in neuroimaging, where the different parts of the $T_2$ distribution are assumed to arise from the different anatomical compartments in brain tissue, particularly in white matter. This can be used, for instance, to generate a map of the myelin water fraction (MWF) such that areas of demyelination corresponding to the effects of neurodegenerative disorders can be identified \citep{mackay2007myelin}. In particular, it is commonly assumed/modelled that the $T_2$ distribution in white matter contains multiple lobes having well-separated peaks, and that the eventual overlap between the $T_2$ lobes of myelin and the intra/extra axonal space water pools is minimal \citep{mackay1994vivo,whittall1997vivo,vasilescu1978water,menon1991application,menon1992proton}.

\begin{figure*}
		\includegraphics[width=\textwidth]{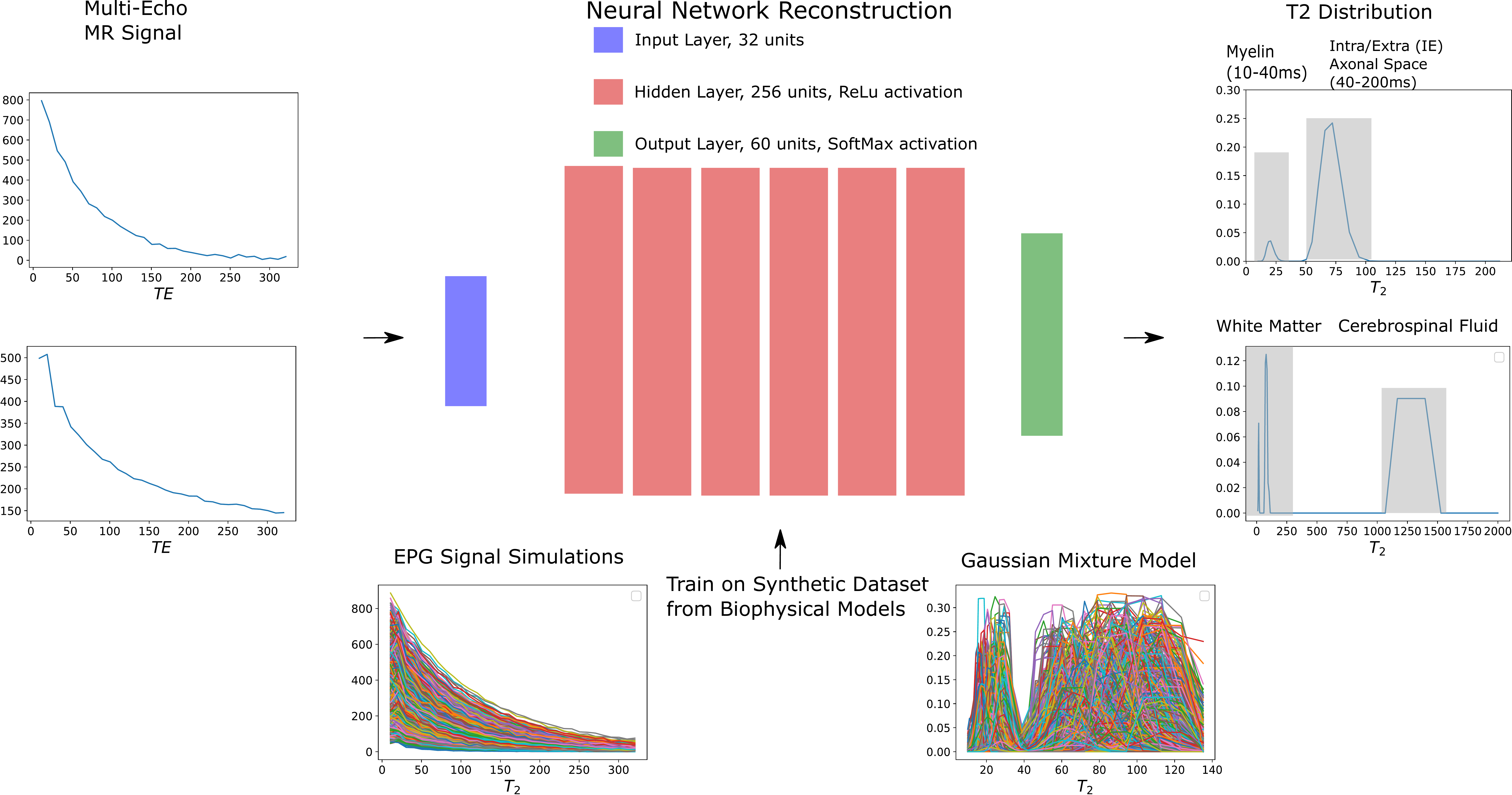}
		\caption{Here we show an overview of our method for multicomponent $T_2$ relaxometry. On the left are example MR signals, on the right are the corresponding $T_2$ distributions: the first distribution is in white matter (WM), where there are assumed to be two lobes: one at a $T_2$ of around 10-40ms corresponding to myelin water and one at a $T_2$ of around 50-120ms corresponding to the intra and extra axonal space. The second distribution includes WM and cerebrospinal fluid (CSF), whose $T_2$ is commonly assumed to be around 1-2s. Our method consists of training a neural network on a synthetic dataset derived from biophysical models to learn the mapping from signal to distribution. At the bottom, we show a small subset of 1000 simulated signals and corresponding $T_2$ distributions from our synthetic training dataset. 
		}
		\label{fig:overview}
\end{figure*}

 \subsection{Related Work}
 In order to estimate $p(T_2)$ from equation (1), two main approaches are generally used: parametric and non-parametric approaches. Parametric approaches rely on \textit{a priori} information on the $T_2$ distribution in brain tissue, particularly white matter, in order to fit the parameters of biophysical models to the MRI signal \citep{raj2014multi,du2007fast,yu2019robust,chatterjee2018multi,akhondi2014t,bjork2016multicomponent}. In these approaches, the MRI signal is modelled as a linear combination of signals from a fixed number of water pools (around 2-3) such as myelin water, the water in the intra-/extra-axonal space, and cerebrospinal fluid:
 \begin{align}
      p(T_2)=\sum_i^n v_i F_i(\mathbf{m}_i,T_2)
 \end{align}

 Here $n$ is the number of water pools assumed, and $F_i,\mathbf{m_i},v_i$ are the probability distribution, parameters of the probability distribution, and volume fraction of the $i$th water pool. A wide variety of parametric distributions (Delta, Gaussian, Truncated Gaussian, Wald, Gamma, Log-Gaussian, Laplacian) are used to model the $T_2$ distributions in these pools; however, \citep{raj2014multi} shows that using these different distributions have negligible differences on the corresponding signal when using the same means and variances; they conclude that due to the ill-posedness of the inverse problem, extracting more than general lobular shapes (characterized by the mean and variance) is extremely difficult if not impossible, even at extremely high signal to noise ratios(SNR). 
 The parameters estimated are the water volume fractions and the parameters of the distributions which are done through optimization \citep{chatterjee2018multi,bjork2016multicomponent} or Monte Carlo methods \citep{prange2009quantifying,yu2019robust}. To stabilize the fitting and to use prior information on the compartments, constraints are enforced on the parameters. For instance, usually bounds are placed on the parameters such as the mean $T_2$ of each compartment; e.g. the mean $T_2$ of myelin water is typically bounded between 10 and 40ms, and the mean $T_2$ of CSF is typically assumed to be greater than 1s. Some works, such as \citep{chatterjee2018multi}, go even further and fix the mean or standard deviations of the probability distributions of some compartments to predetermined values. While parametric estimations are generally stable and histologically validated, they are usually computationally expensive and restricted by the biophysical model used; the number of compartments needs to be fixed for each voxel before fitting. Further, we note that the \textit{a priori} information used in the parametric approaches i.e. the assumption of lobular structure, bounds on the parameters of the distribution, etc. comes from historical evidence, where studies used \textbf{non-parametric} methods to estimate the $T_2$ distributions and assigned lobes in their reconstructions to different water pools \citep{alonso2015mri}. 
 
 In contrast, non-parametric approaches do not make \textit{a priori} assumptions on the data, such as the number of compartments. This is relevant for studying abnormal brain tissue, where compartments not considered in standard biophysical models might be present \citep{mackay2007myelin}. In addition, they generally require orders of magnitude less computation time than parametric methods. Non-parametric methods discretize equation (1) as a product of a dictionary matrix and a discretized $T_2$ distribution and solve directly for the discretized $T_2$ distribution \citep{whittall1997vivo,prasloski2012applications}  using
non-negative least squares (NNLS) algorithms \citep{lawson1995solving}. The $T_2$ distribution, $p(T_2)$, is recovered by solving an inverse problem \citep{whittall1997vivo,prasloski2012applications}. First, given discretized ranges of flip angle ($\alpha$) values and $T_2$ values, a dictionary $D_{\alpha}$ of $T_2$ decay signals is constructed for each $\alpha$ value through the EPG formalism. $D_{\alpha}$ is a matrix where the columns are the simulated MRI signals over a range of $T_2$ values. Given a flip angle $\alpha$, the corresponding dictionary $D_{\alpha}$, and the MRI signal $\mathbf{s}$, the following optimization problem is solved
\begin{align}
    \argmin_{\mathbf{p}\geq 0} &\|D_{\alpha}\mathbf{p}-\mathbf{s}\|_2^2 + \lambda \Phi(\mathbf{p}) 
\end{align}
where $\Phi$ is a regularization function with parameter $\lambda$, and $\mathbf{p}$ is the discretized, un-normalized $T_2$ distribution to be estimated. The flip angle corresponding to $\mathbf{s}$ is chosen by solving the above problem (with $\lambda=0$) for multiple values of $\alpha$ and taking the value which corresponds to the least fitting error \citep{prasloski2012applications}. 
Two common choices for $\Phi(\mathbf{p})$ are
\begin{itemize}
    \item $\Phi(\mathbf{p})=\|\mathbf{p}\|_2^2$, called Tikhonov regularization.
    \item $\Phi(\mathbf{p})=\|\mathbf{L}\mathbf{p}\|_2^2$, where $\mathbf{L}$ is a finite difference approximation of the Laplacian operator. This is called Laplacian regularization.
\end{itemize}
These choices are used in order to promote increased conditioning of the problem and the smoothness of the resulting distribution \citep{kroeker1986analysis}. 
Without regularization, solutions to Eq. (3) are vulnerable to noise and usually produce inaccurate solutions that overfit the signal with e.g. false positive peaks, etc. A common heuristic for selecting $\lambda$ is to accept $\lambda$ such that the signal fitting error is approximately 1.02-1.025 times greater than the error from NNLS with no regularization \citep{laule2006myelin}. However, it is known that regularization can introduce undesirable bias to the reconstructed signals, e.g. over-smoothing. 
In particular, regularization can contradict the expectation of disparate lobes in the distribution corresponding to disparate tissues in the same voxel (e.g. myelin and intra/extra axonal space water), particularly at lower SNRs. For example, at low SNRs, the myelin water lobe can become completely over-smoothed, for an example see Fig. 1 in the Supplementary Material.


Once the $T_2$ distribution is recovered, generating parameters of interest such as volume fractions of the water pools in the voxel require either a distribution where distinct lobes can be assigned to distinct compartments (such as in the right side of Fig. \ref{fig:overview}) or \textit{a priori} information. After examining distributions reconstructed from experimental scans, the different lobes of the distributions (if distinct lobes are present) are assigned to different water pools based on theoretical and experimental grounds \citep{mackay2007myelin}. From the mean and standard deviation of these lobes, bounds are derived from the $T_2$ distributions for each water pool. Then water volume fractions for each pool are calculated by integrating the probability distribution between the bounds of the $T_2$ for each pool.  For instance, at 3T the myelin water fraction (MWF) is usually computed as
\begin{align}
    MWF=\frac{\int_{T_2=10ms}^{T_2=40ms} \mathbf{p}(T_2) \mathrm{d}T_2}{\int_{T_2=10ms}^{T_2=2000ms} \mathbf{p}(T_2) \mathrm{d}T_2},
\end{align}
where the bounds 10-40ms were obtained from the myelin water lobe in NNLS reconstructions in past papers \citep{alonso2015mri}.

We note previous studies found that both parametric and non-parametric methods require a high signal-to-noise ratio (SNR) to detect different components in the $T_2$ distribution \citep{Graham1996,Andrews2005pools,wiggerman}. For a clinically achievable SNR=100, more than 5$\%$ of the voxels were incorrectly estimated to have no myelin water component, and the percentage raised to 12$\%$ for SNR=50 \citep{Kumar2012bayes}. Similar results were reported in \citep{Raj2014spatial}, where the myelin water component was not found in human brain regions located in myelinated areas of the frontal and lateral projections fibers. In addition, \cite{wiggerman} found that in synthetic studies, NNLS with Tikhonov Regularization tends to underestimate the true MWF value in the range of 0.3 to 4 percent at SNR 1000, with the problem worsening at lower SNRs; for reference, the MWF is assumed to be in the range of 0-30 percent in normal appearing white matter.

Recently, \citep{lee2019artificial,LIU2020116551} have both proposed to augment non-parametric approaches with machine learning in order to speed up the computation time. As training data, they acquired brain scans in several subjects \textit{in vivo} using a 3D multiple echo gradient and spin echo sequence with 32 echoes \citep{prasloski2012rapid}. They then ran regularized NNLS reconstructions on the data and obtained the probability distributions and MWF for each voxel. \cite{LIU2020116551} trained a multi-layer perceptron (MLP) to take as input the raw data, and output the MWF, using the \textit{in vivo} NNLS reconstructions as ground truth. \cite{lee2019artificial} trained MLPs to reconstruct the MWF as well as the probability distributions from the raw echo data, using the \textit{in vivo} NNLS reconstructions as ground truth. These approaches have the advantage of reconstructing regularized NNLS solutions for the whole brain in under a minute, a fraction of the time required using the standard NNLS algorithm. However, as their ground truth is the regularized NNLS solution, their method inherits all the problems of NNLS. Further, by training on data acquired from specific MRI machines using a specific sequence, there is the problem of generalizing to different machines and different sequences. Both would require new acquisitions as well as additional training time. 


In summary, parametric methods implicitly regularize and stabilize the problem by using biophysical models and prior knowledge to constrain the space of $T_2$ distributions. However, the resulting optimization problems to be solved are significantly more costly than those of non-parametric methods, with an additional loss of flexibility due to imposition of the number of compartments and other details of the model. 
Non-parametric solutions are fast, but also ill-posed and highly susceptible to noise; hence, regularization is necessary, with the concomitant drawbacks of over-smoothing and sparsity of the reconstructed distributions, particularly at clinically achievable SNRs for sequences with high spatial resolution. Further, the extraction of parameters of interest such as the MWF is theoretically based on assuming a lobular structure of the reconstructed distribution, which is often not the case in middling to high levels of noise. 

\subsection{Contributions}

In this paper, we propose a new method for multi-component $T_2$ relaxometry in brain tissue. In Fig. \ref{fig:overview}, we show the overview of our proposed method as well as a prototypical $T_2$ distribution in white matter, composed of the myelin water lobe and the lobe corresponding to the water in the intra/extra axonal space; in addition, we show the corresponding MRI signal. We propose to combine machine learning and aspects of parametric and non-parametric approaches to the reconstruction of $T_2$ distributions from multi-echo $T_2$ data. We do this by creating a synthetic dataset derived from biophysical models and training a multi-layer perceptron (MLP) \citep{rosenblatt1958perceptron} on this dataset to take as input the MRI signal and directly output the associated $T_2$ distribution. We call our method Model-Informed Machine Learning (MIML). Our main contributions are as follows:
\begin{itemize}
    \item Construction of an extensive synthetic dataset that we construct purely from simulations guided by biophysical models, which we use for training the MLP.
    \item Introduction of a robust loss function for the network to recover the $T_2$ distribution consisting of a combination of the mean squared error and the Wasserstein-1 Distance \citep{villani2009wasserstein}. We show that training with the Wasserstein distance significantly increases the accuracy of MWF estimates on a realistic, synthetic case, compared to training with solely a mean squared error (MSE) loss function. 
    \item Rigorous and extensive evaluation of our method and previous work in non-parametric and parametric approaches, on synthetic and real datasets (\textit{ex vivo}, \textit{in vivo}, healthy, pathological). We show that our method outperforms other methods in terms of accuracy, plausibility, and robustness of the reconstructed distributions and MWF maps as well as lesion visualization. 
\end{itemize}

\section{Methods}
Our method for reconstructing $T_2$ distributions from MRI data is based on a MLP which is trained to learn a map directly from MRI signals with a 32 echo acquisition scheme to the corresponding $T_2$ distribution, as is the result in non-parametric methods. To reduce the inherent ill-posedness of this problem, the training is conducted on a synthetic dataset of pairs of MRI signals and $T_2$ distributions which we constructed using EPG simulations and is informed by biophysical models and realistic values for the parameters of interest, such as the range of $T_2$s for different water pools, taken from the literature. This implicitly constrains the space of possible $T_2$ distributions (as in parametric approaches). We show an overview of our method in Fig. \ref{fig:overview}.

\subsection{Synthetic Dataset Generation}
To generate the synthetic $T_2$ distributions, we start from standard biophysical models for the brain \citep{whittall1997vivo}. Concretely, we model
each distribution as a mixture of Gaussians, where each component corresponds to a different water pool (e.g. myelin water, intra/extra axonal space water). 

\begin{align}
    p(T_2)=\sum_i \frac{v_i}{\sigma \sqrt{2 \pi}}\exp{(\frac{-(T_2-\mu_i)^2}{2\sigma_i^2})}\\
    v_i \in [0,1], \sum_i v_i =1
\end{align}
Here $v_i$ is the volume fraction of the $i$th water pool, and $\mu_i,\sigma_i$ are the mean and standard deviation of the $T_2$ distribution of the $i$th water pool. We justify our choice of modelling using Gaussians by noting that \citep{raj2014multi} found that modelling the $T_2$ distributions using a variety of different distributions including the Gaussian distribution had insignificant differences in parametric methods. For the water pools in brain tissue, most models consider white matter (which in turn contains the myelin and intra/extra axonal water pools) and  cerebrospinal fluid (CSF) \citep{raj2014multi,du2007fast,yu2019robust,chatterjee2018multi,akhondi2014t}. However, brain pathologies can result in $T_2$ distributions different from those of these commonly used compartments. 
In our dataset, we divide $T_2$ distributions in the brain into seven cases, each with a characteristic mixture of water pools.
\begin{itemize}
  \item White matter (WM)
  \item Cerebrospinal fluid (CSF)
  \item Gray matter (GM)
  \item Mixture of WM and CSF
  \item Mixture of WM and GM
  \item Mixture of CSF and GM
  \item Pathology
\end{itemize}
We further split WM into two constituent components
\begin{itemize}
    \item Myelin 
    \item Intra/Extra-axonal Space (IES)
\end{itemize}
As we model each water pool's $T_2$ distribution with a Gaussian distribution, we need to specify the mean ($\mu$) and the standard deviation ($\sigma$). To generate a large variety of signals, we randomly select the means and standard deviations within a range characteristic of the water pool. Concretely, in Table \ref{tab:water_pool_parameters}, we show the range of the means and standard deviations we use for each water pool. For example, for the case of CSF and GM, the Gaussian mixture would be a randomly weighted sum of two Gaussian distributions, with parameters drawn from the CSF range and the GM range, respectively. 
\begin{table*}[]
\centering
\textbf{ Range of Mean and Standard Deviation for Simulated Water Pools}
\vspace{2mm}

\centering
\vspace{2mm}
\begin{tabular}{lll}
\hline
\multicolumn{1}{|l|}{Water Pool} & \multicolumn{1}{l|}{Range of Mean $T_2$ ($\mu$)} & \multicolumn{1}{l|}{Range of Std. of $T_2$ ($\sigma$)} \\ \hline
Myelin                           & 15-30ms                                          & 0.1-5ms                                                \\
Intra/Extra Axonal Space (IES)      & 50-120ms                                         & 0.1-12ms                                               \\
GM                               & 60-300ms                                         & 0.1-12ms                                               \\
Pathology                        & 300-1000ms                                       & 0.1-5ms                                                \\
CSF                              & 1000-2000ms                                      & 0.1-5ms                                               
\end{tabular}
\vspace{2mm}
\caption{Here we show the ranges for the possible mean ($\mu$) and the standard deviations ($\sigma$) used for the Gaussian parameters of the different water pools.  }
\label{tab:water_pool_parameters}
\end{table*}

These cases account for partial-volume effects from the mixing of water pools within a compartment. We note that new cases can easily be incorporated into our approach. We set the mean values in line with those reported in the literature \citep{mackay1994vivo,laule2007long,wansapura1999nmr,alonso2015mri}. We included an expansive range for the standard deviations, ensuring that our dataset has both sparse, intermediate, and wide $T_2$ distributions in order not to bias our dataset towards any extreme. We note that as there is a small quantity of myelin in gray matter, the gray matter pool is composed primarily of the GM component in Table \ref{tab:water_pool_parameters} as well as the myelin water component which is constrained to have a random $v_i$ between 0 and 5 percent. 

For each $T_2$ distribution, we use the EPG formalism to simulate the corresponding signal from an acquisition based on acquiring 32 echos with around 10ms spacing between each echo. In the real data we use for our evaluation, three slightly different echo times are used; the \textit{in vivo} scans of healthy subjects use an echo train of 10.68ms, 21.26ms, ... 341.76ms, the \textit{in vivo} scan of the subject with pathology uses an echo train of 10.36 ms, 20.72 ms, ..., 331.52 ms, and the \textit{ex vivo} scan uses an echo train of 10ms, 20ms, ... 320ms. In the following, we describe our procedure with a single, fixed echo train: for the evaluation, we generated three datasets, one for each echo train. We note that alternative sequences with different numbers of echoes/different spacings can be accommodated by generating a new dataset. 

We generate 200,000 $T_2$ distribution variations per case by sampling ($v_i$) and ($\mu_i,\sigma_i)$  randomly from flat Dirichlet and uniform distributions, for a total of 1.4 million distributions. The corresponding signals are generated using the EPG formalism, and we randomly vary the flip angle ($\alpha$) of the acquisition for each signal between 90 and 180 \degree\ so that our method learns to account for different flip angles automatically, rather than having to first estimate the flip angle as in non-parametric methods. The ground truth distributions were generated on high-resolution grids for the signal generation, then downsampled to the $T_2$ discretization used in the non-parametric approaches to allow for direct comparison. We use a discretization of 60 $T_2$'s logarithmically spaced from 10ms to 2000ms for our $T_2$ distributions. 

As outlined in the related work, the SNR of the signals is a crucial aspect of the reconstruction and hence the dataset generation. We define SNR with respect to the first echo of the signal sequence. From previous studies \citep{wiggerman,mackay1994vivo}, it is known that NNLS methods, perform well in the high-SNR regime (on the order of 1000). However, clinical scans with high spatial resolutions will rarely meet this SNR requirement; in the real scans of healthy subjects we use in our evaluation, we estimate a mean SNR on the order of 100. In order to make our method robust to the realistically low SNR regime, in training we randomly vary the SNRs of the signals between 80 and 200 in order to cover the potential SNR range of the voxels. We use a Rician noise model to add noise to the signals. In our evaluation, we show that training on this SNR range results in robustness to a wide range of SNRs (40-1000) on synthetic data. The data generation for all cases (1.4 million signal/distribution pairs) took approximately 24 hours on a cluster using parallelization on 46 threads. 

Using the synthetic datasets described, we train a MLP to map the MRI signal to the corresponding $T_2$ distribution.

\subsection{Mapping the MR Signal to the $T_2$ Distribution}
\subsubsection{Architecture}
Our network is composed of 6 hidden layers with 256 neurons per layer and an output layer with 60 units, corresponding to the size of the discretization of the distributions we use. The hidden layers use a ReLu function as the activation function, while the output layer uses a SoftMax activation function since the output should be the $T_2$ distribution. The input to the network is a vector with 32 elements corresponding to the 32 echos of the standard acquisition sequence. We note that we normalize the input by the magnitude of the first echo before feeding it to the network. To select the structure of the network, we trained 12 networks where we varied the number of hidden layers (3-6) and the number of neurons per layer (64,128,256,512). We selected 6 hidden layers and 256 neurons as this configuration had the lowest validation loss at the end of training; however, we note that the validation loss was not significantly different between the configurations.

\subsubsection{Loss Function}
Let $(\mathbf{x},\mathbf{p_x})$ denote the normalized MRI signal and the corresponding $T_2$ distribution. Let $\Phi(\cdot,\mathbf{\theta})$ denote the multi-layer perceptron function with parameters $\mathbf{\theta}$, with $\Phi(\mathbf{x},\mathbf{\theta})$ the predicted distribution. Given a batch of training samples of size $n$, the cost function we use to train $\Phi$ is
\begin{align}
    L(\theta)=\frac{1}{n} \sum_i^n  \lambda\|\mathbf{p_x}-\Phi(\mathbf{x},\mathbf{\theta})\|_2^2 + W_1(\mathbf{p_x},\Phi(\mathbf{x},\mathbf{\theta}))
\end{align}
where the first term corresponds to the squared $L_2$ norm (MSE loss) and the second term corresponds to the Wasserstein-1 distance on probability distributions \citep{villani2009wasserstein}. We set $\lambda$ to give approximately equal numerical weight to both terms in the loss function.  Let $u,v$ denote 1-D probability distributions with cumulative distribution functions $U,V$. Then the Wasserstein-1 Distance is equivalent to the following formulation \citep{ramdas2017wasserstein} \begin{align}
    W_1(u,v)=\int_{-\infty}^{\infty} |U-V| 
\end{align}
In this formulation, the Wasserstein distance can be efficiently computed on GPU using the cumulative sum function. 
The Wasserstein-1 distance is an appropriate metric to judge reconstruction quality in our application of $T_2$ distribution recovery as it correctly penalizes deviations from the ground truth distribution in relation to the location of the lobes in contrast to other losses such as MSE or Kullback-Liebler (KL) divergence. In particular, given two non-overlapping lobes, if the lobes are moved toward each other (but still do not overlap), the Wasserstein Distance will decrease significantly while the MSE and the KL Divergence will not change. An example is presented in Fig 2. in the Supplementary material.  


Using the Wasserstein distance helps us to avoid, for example, cases where the location of lobes in the distribution could be arbitrarily placed with a similar loss if other metrics are used. We note that training with either MSE loss or Wasserstein-1 distance exclusively leads to suboptimal results, due to increased Wasserstein-1 distance in the first case and unstable reconstructions in the second case. We find that training with a combination of these results worked optimally; we further show in our evaluation that adding the Wasserstein-1 distance improves the accuracy of MWF estimation in realistic cases in comparison to training exclusively with MSE loss. 

\subsubsection{Implementation Details}
We used TensorFlow 2.0 \citep{tensorflow2015-whitepaper} on Python 3.6 \citep{van2000python} with an Nvidia GTX 2070 laptop GPU for constructing and training the network. For each case, we use 80 percent of the generated data for training, corresponding to a total of 1,120,000 signal/distribution pairs. We reserve 10 percent of the dataset as the validation set and the remaining 10 percent as the test set in our evaluation on synthetic data.  We use the Adam optimizer \citep{kingma2014adam} with a learning rate of 5e-4 and a batch size of 2000. We trained for 30 epochs, where we stopped the training based on the validation loss no longer decreasing for 5 epochs. This training took approximately 15 minutes to complete, showing the feasibility, given a large database of signals, to retrain models specific to given sequences, etc.

\section{Evaluation}
We perform reconstructions of the $T_2$ distributions from synthetic and real data using the following methods:
\begin{itemize}
    \item Our proposed method, MIML,trained on signals with SNR 80-200 and the appropriate sequence of echoes. 
    \item NNLS with Tikhonov regularization (NNLS-T) \citep{mackay1994vivo}
    \item NNLS with Laplacian regularization (NNLS-L) \citep{prasloski2012applications}
    \item Gaussian Mixture Fitting  (GMF)
\end{itemize}
Both NNLS methods were implemented in-house in Python with full parallelization, and we use a standard method of selecting the regularization parameter \citep{prasloski2012applications} by keeping the signal fitting error close to 1.025 times the signal fitting error obtained using NNLS without regularization. GMF is our implementation of a parametric approach, where we fit a Gaussian mixture model with three compartments (myelin water, IES water, CSF), extracting the volume fractions, the means/standard deviations of the $T_2$ of each compartment, and the overall normalization factor. We model as follows:
\begin{align}
     p(T_2)=\sum_{i=1}^3 v_i \mathcal{N}(\mu_i,\sigma_i,T_2).
\end{align}
We set bounds on the means/standard deviations according to the bounds used for generating the dataset for MIML. As simultaneously fitting the flip angle resulted in severe instability, we fix the flip angle in the Gaussian mixture fitting for each voxel to that calculated using a standard method used with NNLS \citep{prasloski2012applications}. 
We used the least squares optimization function in the Python library Scipy \citep{2020SciPy-NMeth} to fit the signals to the Gaussian model. 

\subsection{Synthetic Data}
\subsubsection{Test Split of Synthetic Dataset}
We show reconstructions on the test split of the synthetic dataset we generated using the acquisition sequence of 10.68ms, 20.68ms, ... 320.68ms. We show results over an SNR range from 40 to 1000 (40,80,150,200,400,1000). We compare the methods using the mean MSE and Wasserstein Distances with respect to the ground truth.

\subsubsection{Realistic Synthetic Case in WM}
MWF mapping is a crucial application of $T_2$ relaxometry. In order to analyze the robustness and performance of our approach in a realistic case in WM, we show reconstructions on the following model of the distribution in a white matter voxel, with one lobe for myelin water and one lobe for IES water.
\begin{align}
    p(T_2)=v_m* InvGamma(\mu_m,\sigma_m)+v_{IE}* InvGamma(\mu_{IE},\sigma_{IE})
\end{align}
where we fix the values of the parameters to realistic values in line with those reported in the literature \citep{alonso2015mri,mackay2007myelin}: $v_m$=0.15, $v_{IE}$=0.85, $\mu_m$= 20ms, $\mu_{IE}$= 70ms, $\sigma_m$=2.5ms, $\sigma_{IE}$=6ms.
We use the inverse Gamma distribution to create the ground truth distribution to test the robustness of our method to changes
in the assumed biophysical model. To study robustness to noise, we vary the SNR on the corresponding synthetic MRI signal from 40 to 1000, as in the test split. We generate 1000 realizations of noisy signals per SNR used. Further, we also show numerical results using our method \textbf{without} using the Wasserstein Distance in the loss function. \textbf{We refer to this variant as MIML'}. We compare the methods using the mean MSE, Wasserstein Distance, and estimated MWF with respect to the ground truth.

\subsection{Real Data}
 As there is no ground truth for the $T_2$ distributions in real data, we evaluate the methods as in the literature by examining the MWF maps/comparing to anatomical scans or correlation to histology, the plausibility of the $T_2$ distributions, maps of the mean $T_2$ in the 50-200ms range, etc. We also report the mean SNR for each dataset, calculated in the same manner as in \cite{wiggerman}, where the first echo of the signals is divided by the standard deviations of the residuals from the NNLS-T reconstruction.
\subsubsection{\textit{Ex Vivo} Data}
We show reconstructions from a Multi Echo Spin Echo (MESE) scan from the White Matter Microscopy Database \cite{Cohen} with 32 echoes (starting from 10ms with 10ms spacing), with a TR of 3s and 8-fold averaging, of a single, cervical slice of a dog's spinal cord acquired \textit{ex vivo} with an Agilent 7T animal scanner \citep{bib:Vuong:2017}. Five days before scanning, the spinal cord (perfused and post-fixed with paraformaldehyde 4) was extracted and washed in Phosphate-buffered saline (PBS) solution. After scanning, the spinal cord was osmified for two hours, embedded in EMbed 812 Resin, cut using a microtome, and polished. A scanning electron microscope (Low-angle backscattered electron mode) (JEOL 7600F) was used to image an entire slice of the spinal cord at a resolution of 0.26 micrometers per pixel. Using this histology image, we construct a histological map of the fraction of myelin in each voxel using a deep learning segmentation tool called Axon Deepseg \citep{zaimi2018axondeepseg}. We then register this histological map to the MRI space. We note that this histological map is the area fraction of the myelin segmented, not a map of the MWF. However, assuming that the area fraction of myelin in a voxel scales with the amount of myelin water, the two maps should be linearly correlated.  We conduct a correlation analysis between the histological map and the MWF maps produced from the different methods. The estimated SNR on this slice is 784. 

\subsubsection{Healthy Subjects}
We show reconstructions from high-resolution human brain scans acquired from 4 healthy controls using a 3T MRI scanner  (MAGNETOM Prisma, Siemens Healthcare, Erlangen, Germany) located at CHUV Hospital (Lausanne, Switzerland), with a standard   64-channel head/neck coil. The dataset was collected using a 3D multi-echo gradient and spin-echo (GRASE) sequence accelerated with CAIPIRINHA \citep{Piredda} with the following parameters: matrix-size=144x126; voxel-size = 1.6x1.6x1.6mm3; $\Delta$TE/N-echoes/TR = 10.68ms/32/1s; prescribed FA =180\degree; number-of-slices = 84; CAIPIRINHA acceleration factor = 3x2; number of averages = 1; acquisition time=10:30min. Each subject was also scanned using an MPRAGE sequence for whole-brain $T_1$-weighted imaging \citep{brant1992mp}. 
To test the repeatability of the reconstructions, the healthy controls were scanned twice over two consecutive scanning sessions (scan-rescan scenario). We compare the MWF maps and the $T_2$ distributions produced from each method, show the coefficient of variability of the MWF in regions of interest (ROI) in WM, and conduct a study of the reproducibility of each method. The data for these subjects have an estimated mean SNR of 128. 
\subsubsection{MS Subject}
We show  reconstructions on a high-resolution human brain scan of a patient with relapsing-remitting multiple sclerosis, scanned using a 3T MRI scanner (MAGNETOM Prisma, Siemens Healthcare, Erlangen, Germany) located at the University Hospital of Basel (Basel, Switzerland) with a standard 32-channel head coil. In this case, MET2 data was collected using the previously described GRASE sequence for the healthy subjects, albeit with a starting echo time of 10.36ms and lower spatial resolution (voxel-size=1.8x1.8x1.8mm3) to accelerate the scan. In addition, a FLAIR \citep{de1992mr} scan was acquired. A probabilistic lesion mask was generated by first using a convolutional neural network (CNN) trained to segment WM lesions \citep{la2020multiple} on FLAIR images with subsequent manual correction by an expert. The FLAIR image/lesion mask were then registered to the multi-echo $T_2$ space. We use a threshold of 0.5 to denote a voxel as lesional. We analyze maps of the geometric mean $T_2$ in the range 50-200ms and MWF maps to study the MS lesions as in \citep{levesque2010quantitative}. We also compare the correspondence of these maps to the lesion masks. In addition, we compare the $T_2$ distributions produced from each method in both normal-appearing tissue and the lesions. The estimated SNR of this scan is 112. 
 
\section{Results}
\subsection{Synthetic Data}
\subsubsection{Test Split of Synthetic Dataset}
\begin{figure*}[h]
		\includegraphics[width=\textwidth]{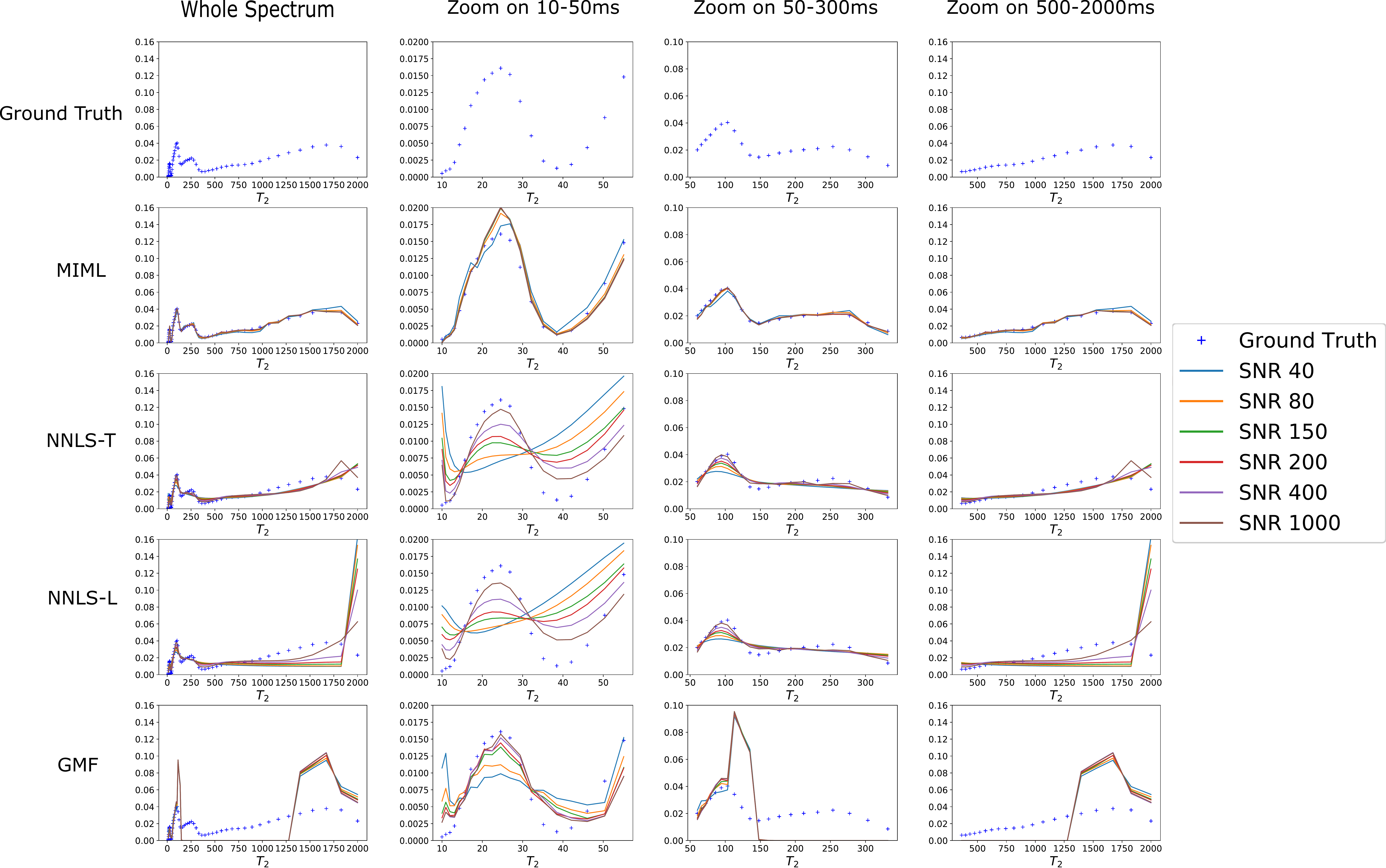}
		\caption{For better visualization, we show plots of the mean distribution over all of the ground truth distributions in the test split, as well as the mean reconstructed distribution over the test split from each method. We zoom in on the $T_2$ ranges 10-50ms, 50-400ms, 400-2000ms to show the average performance in the different cases (WM, CSF, etc.). Our method produces the most robust and accurate reconstructions with respect to changing SNR and the ground truth distributions respectively. All other methods require high SNRs (400-1000) for plausible distributions which still retain distortions, particularly in the $T_2$ range 10-50ms, which is that of the myelin water. We note that the poor performance of GMF outside the $T_2$ range 10-120ms, and 1000-2000ms is due to model mismatch; GMF is valid only for the WM and WM+CSF cases. 
		}
		\label{fig:synthetic_test}
\end{figure*}

 In order to visualize the average performance over the test split, in Fig. \ref{fig:synthetic_test} we plot the mean distribution over all the ground truth distributions in the test split. In addition, we show the mean reconstructed distributions over the test split from the methods we compare. We also show plots zooming in on the different $T_2$ regions for better visualization, as the logarithmic set of $T_2$ values used for the reconstruction makes resolving the lower $T_2$s somewhat difficult. We can see that our method performs robustly and consistently across the whole SNR range, providing the best conformity to the ground truth distributions over the entire range of $T_2$s. In contrast, NNLS with Tikhonov and Laplacian regularization both require SNR 1000 in order to generate a plausible distribution in the $T_2$ range $10-50$ms, with SNRs below this resulting in highly over-smoothed distributions. Further, even at high SNRs, both methods have over-smoothing in the $T_2$ range $50-2000$ms. For the Gaussian mixture fitting, we note that only the cases of WM and WM + CSF correspond to the model used, as it is necessary to fix the number of compartments beforehand. Therefore, the relevant $T_2$ ranges to examine are $10-120,1000-2000ms$. We can see that high SNRs (200-1000) are required for plausible distributions with respect to the ground truth, with remaining distortions at low $T_2$ values. In addition, in Tables \ref{tab:synthetic_test_mse} and \ref{table:synthetic_test_wass}, we show the mean and standard deviations of the MSE and the Wasserstein Distance between the ground truth distributions and the reconstructed distributions from the different methods over the SNR range. As the model used in GMF only applies to WM and WM+CSF, we show the results  over the whole test set as well as over just the WM and WM+CSF cases in the test set. 

\begin{table*}[]
\centering
\textbf{ Mean and Standard Deviation of MSE (Whole Dataset/WM,WM+CSF)}
\vspace{2mm}

\centering
\vspace{2mm}
\begin{tabular}{llllll}

    SNR     & MIML          & NNLS-T                             & NNLS-L        & GMF           &  \\
 40   & (\textbf{0.098},0.115,)/(\textbf{0.056},0.045) & (0.134,\textbf{0.085})/(0.074,\textbf{0.038})      & (0.223,0.229)/(0.091,0.077) & (0.277,0.2)/(0.133,0.085)   &  \\
 80   & (\textbf{0.056},\textbf{0.076})/(\textbf{0.038},\textbf{0.031}) & (0.117,0.083)/(0.062,0.038)       & (0.201,0.218)/(0.082,0.074) & (0.26,0.207)/(0.105,0.075)  &  \\
 150  & (\textbf{0.039},\textbf{0.055})/(\textbf{0.032},\textbf{0.027}) & (0.102,0.083)/(0.052,0.035)       & (0.173,0.199)/(0.072,0.072) & (0.248,0.214)/(0.087,0.071) &  \\
 200  & (\textbf{0.036},\textbf{0.053})/(\textbf{0.031},\textbf{0.026}) & (0.097,0.081)/(0.049,0.034)       & (0.157,0.182)/(0.067,0.07) & (0.244,0.215)/(0.075,0.064) &  \\
 400  & (\textbf{0.034},\textbf{0.049})/(\textbf{0.029},\textbf{0.026}) & (0.088,0.081)/(0.045,0.036)       & (0.124,0.149)/(0.055,0.063) & (0.238,0.219)/(0.067,0.066) &  \\
 1000 & (\textbf{0.033},\textbf{0.048})/(\textbf{0.029},\textbf{0.026}) & (0.085,0.085)/(0.042,0.037)       & (0.091,0.106)/(0.041,0.051) & (0.233,0.223)/(0.056,0.061) & 
\end{tabular}
\vspace{2mm}
\caption{Here we show the mean and the standard deviation of the MSE over the whole test set and just the WM,WM+CSF cases between the ground truth distributions and the reconstructions for each method over a range of SNR values. We see that MIML produces significantly lower values for the mean and the standard deviation of the MSE than those of the other methods over the majority of the SNR range over both the whole dataset and just the WM,WM+CSF cases. The performance of GMF significantly improves on the WM,WM+CSF cases as these fit the assumed model in this method. }
\label{tab:synthetic_test_mse}
\end{table*}




\begin{table*}[]
\centering
\textbf{ Mean and Standard Deviation of Wasserstein Distance (Whole/WM,WM+CSF)}
\vspace{2mm}

\centering
\vspace{2mm}
\begin{tabular}{llllll}
         & MIML         & NNLS-T        & NNLS-L        & GMF           &  \\
SNR 40   & (\textbf{92.2},132.4)/(\textbf{58.3},\textbf{70.4}) & (139.9,\textbf{131.2})/(76.8,78.3) & (188.1,190.5)/(88.4,98.5) & (261.1,261.0)/(76.4,84.1) &  \\
SNR 80   & (\textbf{58.5},\textbf{92.5})/(\textbf{43.9},\textbf{56.8})  & (116.6,119.2)/(63.0,74.4) & (163.7,174.9)/(77.3,97.6) & (251.0,265.7)/(64.9,83.9) &  \\
SNR 150  & (\textbf{44.4},\textbf{72.2})/(\textbf{40.1},\textbf{53.1})  & (98.2,107.8)/(55.2,68.8)  & (139.9,156.3)/(69.7,91.9) & (245.7,268.7)/(59.1,81.0) &  \\
SNR 200  & (\textbf{42.7},\textbf{70.1})/(\textbf{39.4},\textbf{52.5})  & (92.0,103.1)/(53.8,68.9)  & (128.7,145.2)/(67.6,92.1) & (244.1,269.6)/(55.8,79.7) &  \\
SNR 400  & (\textbf{40.7},\textbf{67.6})/(\textbf{38.7},\textbf{51.9})  & (76.2,90.5)/(49.5,66.1)   & (102.6,120.7)/(60.1,84.2) & (240.1,270.6)/(50.1,74.5) &  \\
SNR 1000 & (\textbf{40.2},\textbf{66.5})/(\textbf{38.5},\textbf{51.7})  & (58.0,73.3)/(42.8,58.6)   & (72.1,91.7)/(48.5,69.8)   & (237.6,272.5)/(45.0,68.8) & 
\end{tabular}
\vspace{2mm}
\caption{Here we show the mean and the standard deviation of the Wasserstein Distance over the whole test set and just the WM,WM+CSF cases between the ground truth distributions and the reconstructions for each method over a range of SNR values. We see that MIML produces significantly lower values for the mean and the standard deviation than those of the other methods over the majority of the SNR range.  The performance of GMF significantly improves on the WM,WM+CSF cases as these fit the assumed model in this method.}
\label{table:synthetic_test_wass}
\end{table*}


For both MSE and Wasserstein Distance, MIML performs the best in terms of the mean value, with comparable or lower standard deviations, across the whole SNR range. As expected, all methods improve with increasing SNR. In addition, we can see that limitations of the GMF model, as it provides competitive results with the other methods only when restricted to the signals from the WM and WM+CSF cases, due to the need to fix the model/number of compartments beforehand.
Overall, we can see that MIML, which is trained on signals with SNR 80 to SNR 200, generalizes well to the test set as well as to SNRs outside the range on which it was trained. From the plots of the mean distributions and the tables of the mean metrics, MIML performs better in distribution reconstruction than the other methods, parametric and non-parametric, across a wide range of SNRs. However, the test set is generated according to the Gaussian mixture model; further, as we randomly generate the ground truth distributions, not all of the ground truth distributions are realistic, though we note that unrealistic distributions in the training can improve the generalizability of MIML. 
\subsubsection{Realistic Synthetic Case in WM}
\begin{figure*}[h]
		\includegraphics[width=\textwidth,height=8cm]{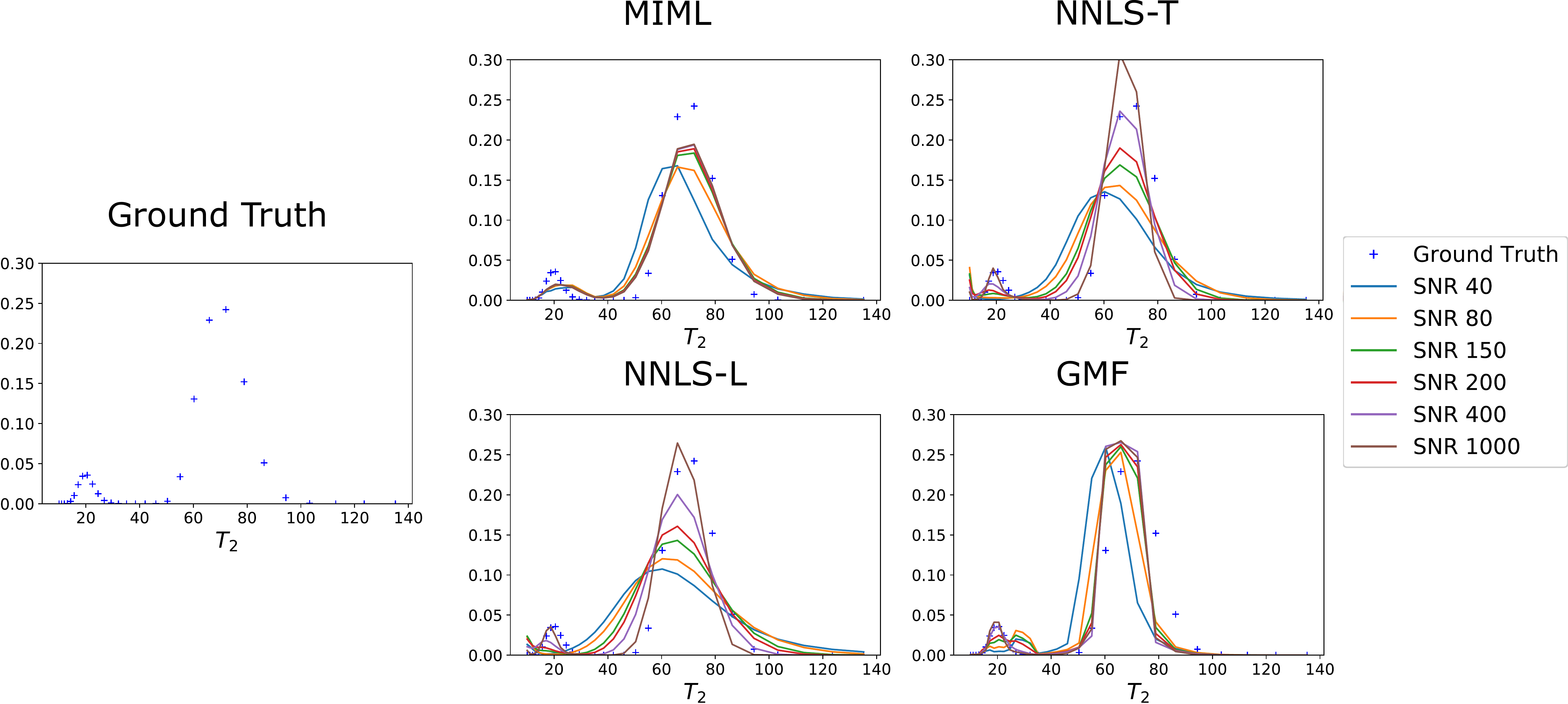}
		\caption{We show plots of the realistic, ground truth distribution, as well as the mean reconstructed distribution from each method. MIML produces the most robust reconstructions with respect to changing SNR, albeit with a consistently over-smoothed myelin water lobe. However, the other methods require high SNRs (1000) to resolve a myelin water lobe (still with distortions) close to the ground truth lobe as well as correct placement of the IE lobe. 
		}
		\label{fig:realistic_case}
\end{figure*}

In Fig. \ref{fig:realistic_case}, we plot the ground truth distribution and the mean reconstructed distributions from each method. We can see that MIML performs robustly and consistently, on average, across the whole SNR range. However, the reconstructed distributions resolve a more spread out myelin water lobe than in the ground truth, even at SNR 1000; this could be due to training on significantly lower SNRs or the model mismatch. At SNRs below 400, NNLS-T and NNLS-L are unable to resolve a myelin water lobe due to over-smoothing as well as a displaced IES lobe; GMF resolves the myelin water lobe, but with a significantly displaced mean. At SNR 1000, NNLS-T and NNLS-L are able to resolve the myelin water lobe accurately, albeit still with a small distortion at $T_2=10$; GMF is able to accurately capture the myelin water lobe at SNR 1000, albeit with a displaced IE lobe. In Tables \ref{tab:realistic_case_mse}, \ref{table:realistic_case_wass}, \ref{table:realistic_case_mwf},  we show the mean and standard deviations of the MSE and the Wasserstein Distance between the ground truth distribution and the reconstructed distributions from the different methods over the SNR range, as well as the mean and standard deviation of the recovered MWF.  With regard to MSE and Wasserstein Distance, MIML performs the best in terms of the mean value, with comparable or lower standard deviations, across the whole SNR range. As expected, MIML' performs similarly to MIML with respect to MSE and significantly worse with respect to Wasserstein Distance, as it is only trained with the MSE loss. With regard to the recovered MWF (obtained by summing from $T_2$ bounds of 10-40ms), we see that MIML performs the best in terms of the mean value, with comparable or lower standard deviations, in the SNR range 80-400. At SNR 40, NNLS-L performs slightly better than MIML in terms of the mean MWF (0.167 vs. 0.132). At SNR 1000, NNLS-T provides a slightly better mean estimate for the MWF than MIML (0.15 vs. 0.146). However, we note that NNLS-T has a significantly higher standard deviation than MIML at SNR 1000 (0.014 vs 0.004). Further, the table is consistent with results in \citep{wiggerman} that the NNLS methods tend to underestimate the MWF. MIML' provides mediocre performance, with low standard deviation values but also with inaccurate mean values. GMF provides the second best performance after MIML in terms of the mean value. Overall, MIML performs accurately and robustly across the whole range of SNRs with respect to the MSE, Wasserstein Distance, and the MWF value, showing the robustness to changing the assumed model as well as the applicability in a realistic case. Other methods perform comparably at high SNR values (SNR 1000), as expected. Finally, comparing the performance of MIML and MIML', we can see that using the Wasserstein Distance in the loss function during training significantly improves the performance of our method in terms of MWF estimation in a realistic case. 

From the results on the synthetic data, we conclude that MIML, even trained on a limited range of SNRs, is able to robustly and accurately reconstruct $T_2$ distributions over a wide range of SNR values. Overall, MIML outperforms all other methods in terms of MSE and Wasserstein Distance with respect to the ground truth, both in terms of the mean and standard deviation of these metrics. Furthermore, from the realistic case, MIML is the most accurate overall method for MWF estimation, showing the applicability to MWF estimation. In addition, we can see the robustness to changes in the assumed model for the $T_2$ distributions, and the importance of including the Wasserstein Distance in the loss function of MIML. In the next section, we show results on real data from \textit{in vivo} and \textit{ex vivo} scans, considering both healthy and pathological cases. 
 
\begin{table*}[h]
\centering
\textbf{ Mean and Standard Deviation of MSE (Realistic Case)}
\vspace{2mm}

\centering
\vspace{2mm}
\begin{tabular}{lllllll}

         & MIML          & MIML'         & NNLS-T        & NNLS-L        & GMF           &  \\
SNR 40   & (\textbf{0.058},0.033) & (0.052,0.029) & (0.088,0.028) & (0.087,\textbf{0.017}) & (0.19,0.11)   &  \\
SNR 80   & (\textbf{0.024},\textbf{0.016}) & (0.026,\textbf{0.016}) & (0.064,0.022) & (0.07,\textbf{0.016})  & (0.101,0.067) &  \\
SNR 150  & (\textbf{0.012},\textbf{0.007}) & (0.014,\textbf{0.007}) & (0.045,0.021) & (0.052,0.016) & (0.066,0.038) &  \\
SNR 200  & (\textbf{0.01},\textbf{0.004})  & (0.011,0.005) & (0.039,0.02)  & (0.044,0.016) & (0.06,0.029)  &  \\
SNR 400  & (\textbf{0.008},\textbf{0.002}) & (0.009,\textbf{0.002}) & (0.032,0.02)  & (0.03,0.013)  & (0.052,0.011) &  \\
SNR 1000 & (\textbf{0.007},\textbf{0.001}) & (0.008,\textbf{0.001}) & (0.042,0.035) & (0.023,0.014) & (0.046,0.013) & 
\end{tabular}
\vspace{2mm}
\caption{Here we show the mean and the standard deviation of the MSE between the ground truth distribution of the realistic case and the reconstructions for each method over a range of SNR values. We see that MIML produces significantly lower values for the mean and the standard deviation of the MSE than those of the other methods over the majority of the SNR range. As expected, MIML' performs almost identically to MIML, as MIML' is trained just with MSE loss. }
\label{tab:realistic_case_mse}
\end{table*}

\begin{table*}[h]
\centering
\textbf{ Mean and Standard Deviation of Wasserstein Distance (Realistic Case)}
\vspace{2mm}

\centering
\vspace{2mm}
\begin{tabular}{lllllll}
         & MIML        & MIML'       & NNLS-T      & NNLS-L      & GMF         &  \\
SNR 40   & (\textbf{16.7},11.5) & (43.1,18.1) & (37.2,13.9) & (32.3,13.2) & (34.8,\textbf{11.4}) &  \\
SNR 80   & (\textbf{5.1},\textbf{2.2})   & (16.8,7.9)  & (15.2,7.9)  & (13.7,6.5)  & (14.9,7.1)  &  \\
SNR 150  & (\textbf{3.3},\textbf{1.0})   & (9.4,3.2)   & (8.2,4.5)   & (8.2,3.4)   & (8.4,3.9)   &  \\
SNR 200  & (\textbf{2.8},\textbf{0.7})   & (8.1,2.1)   & (6.5,3.3)   & (6.6,2.6)   & (7.0,2.8)   &  \\
SNR 400  & (\textbf{2.4},\textbf{0.3})   & (7.2,1.0)   & (4.6,1.7)   & (4.7,1.5)   & (5.4,1.2)   &  \\
SNR 1000 & (\textbf{2.3},\textbf{0.1})   & (7.0,0.3)   & (3.9,0.8)   & (3.6,0.7)   & (4.8,0.5)   & 
\end{tabular}
\vspace{2mm}
\caption{Here we show the mean and the standard deviation of the Wasserstein Distance between the ground truth distribution of the realistic case and the reconstructions for each method over a range of SNR values. We see that MIML produces significantly lower values for the mean and the standard deviation of the Wasserstein Distance than those of the other methods over the majority of the SNR range. Further, MIML' is overall the worst performing method in this metric, with high values for both the mean and standard deviation of the Wasserstein Distance. }
\label{table:realistic_case_wass}
\end{table*}

\begin{table*}[h]
\centering
\textbf{ Mean and Standard Deviation of MWF (Realistic Case)}
\vspace{2mm}

\centering
\vspace{2mm}
\begin{tabular}{llllllll}
         & Ground Truth & MIML          & MIML'         & NNLS-T        & NNLS-L        & GMF           &  \\
SNR 40   & 0.15         & (0.132,0.095) & (0.119,\textbf{0.083}) & (0.129,0.094) & (\textbf{0.167},0.091) & (0.101,0.096) &  \\
SNR 80   & 0.15         & (\textbf{0.153},0.052) & (0.135,\textbf{0.045}) & (0.13,0.056)  & (0.139,0.051) & (0.156,0.073) &  \\
SNR 150  & 0.15         & (\textbf{0.149},0.028) & (0.132,\textbf{0.025}) & (0.13,0.034)  & (0.123,0.028) & (0.166,0.05)  &  \\
SNR 200  & 0.15         & (\textbf{0.148},0.021) & (0.131,\textbf{0.019}) & (0.132,0.03)  & (0.122,0.023) & (0.162,0.041) &  \\
SNR 400  & 0.15         & (\textbf{0.146},\textbf{0.01})  & (0.131,\textbf{0.01})  & (0.14,0.021)  & (0.128,0.017) & (0.158,0.023) &  \\
SNR 1000 & 0.15         & (0.146,\textbf{0.004}) & (0.131,\textbf{0.004}) & (\textbf{0.15},0.014)  & (0.142,0.012) & (0.157,0.014) & 
\end{tabular}
\vspace{2mm}
\caption{Here we show the mean and the standard deviation of the MWF calculated from the reconstructions for each method over a range of SNR values. We show the ground truth value (0.15) in the leftmost column. We see that MIML, overall, produces the most accurate estimation of the MWF in addition to standard deviations on par with or lower than those of other methods over the majority of the SNR range. MIML' performs much worse than MIML in terms of the mean MWF with very similar values for the standard deviation. This highlights the downstream impact of using the Wasserstein Distance in the loss function in our method. }
\label{table:realistic_case_mwf}
\end{table*}

\subsection{Real Data}
\subsubsection{\textit{Ex vivo} Data}
\begin{figure*}
		\includegraphics[width=\textwidth]{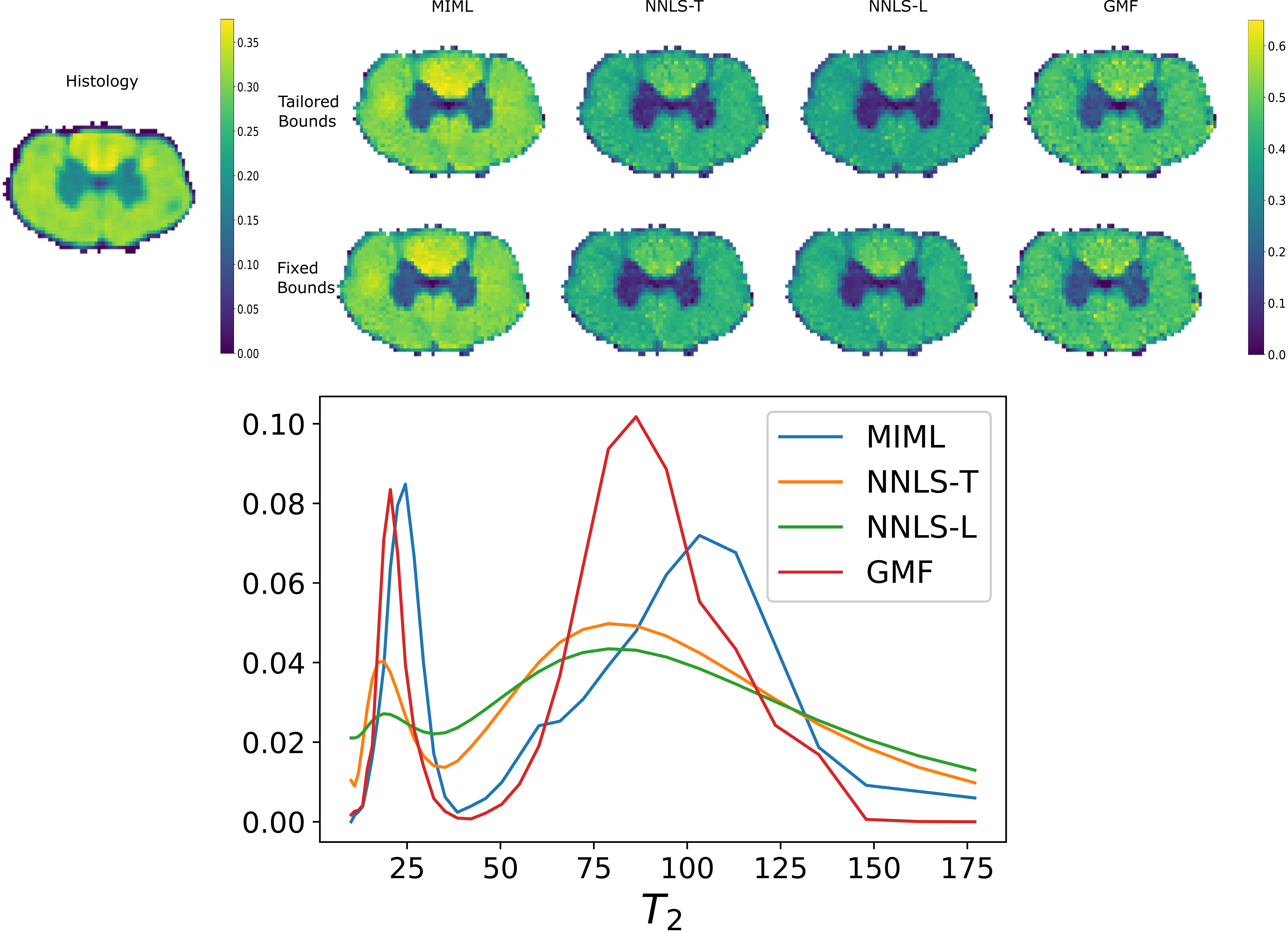}
		\caption{Here we show the MWF maps corresponding to each case for the bounds, the histology map, and the mean distributions over the white matter for each method. We can see that the MIML MWF map from is smoother and corresponds better to the histology map. We can see from the mean distributions that all methods can recover the myelin water and IES water lobe in similar locations; however, the NNLS methods produce over-smoothed lobes in comparison to MIML and GMF. We emphasize that the histology map is a map of the area fraction of the myelin in the voxel, not the MWF. 
		}
		\label{fig:ex_vivo_mwf}
\end{figure*}

We note that in Equation (4), the MWF is obtained by summing from $T_2=10$ms to $T_2=40$ms. This formula, commonly used for acquisitions at 3T, in theory should be adjusted for higher field strengths due to the shortening of $T_2$s \citep{kolind2009myelin,laule2008myelin}. We note that these limits historically derive from assignment of the different lobes in $T_2$ distributions to different water pools e.g. myelin, IE space, etc.\citep{alonso2015mri}. For instance, in \citep{mackay1994vivo}, the authors use NNLS-T on their data (acquired at 1.5T) and found two large $T_2$ lobes, one in the range of 10-50ms and the other in the range of 70-100ms; they then assigned these to myelin water and the IES water respectively. In the following, we restrict our analysis to the white matter, and we will show two versions of MWF maps, with accompanying correlations to histology obtained as follows:
\begin{itemize}
    \item Fixed Limits: Following \citep{mackay1994vivo}, we fix the limits of summation for each method by taking the limits of the myelin water lobe in the mean $T_2$ distribution from using NNLS-T. This corresponds to bounds of  10-35ms.
    \item Tailored Limits: For each method, we set the limits of summation from the limits of the low $T_2$ lobe in the mean $T_2$ distribution from that method. For MIML and NNLS-L this corresponds to bounds of 10-38ms and 10-32ms respectively. 
\end{itemize}
In Table \ref{table:histology_correlations}, we show the Pearson correlations (with accompanying p-values) between the MWF maps for each method and the histology map. We can see that in both cases, the MWF map from MIML has the highest correlation to the histology map. Only the correlation of NNLS-L changes between the two cases, increasing when using the fixed bounds. In Fig \ref{fig:ex_vivo_mwf}, we show the MWF maps corresponding to each case for the bounds, the histology map, and the mean distributions over the white matter for each method. MIML predicts higher values for the MWF than the other methods, particularly the NNLS methods. The MIML MWF map from is smoother/less noisy than the other methods and corresponds better to the histology map. We can see from the mean distributions that all methods are able to recover the myelin water and IES water lobe in similar locations; however, the NNLS methods produce implausible, over-smoothed lobes in comparison to MIML and GMF. 

We note that for all methods, the MWF values are significantly higher than those of the \textit{in vivo} 3T scans we show later. However, this can be attributed to the differences resulting from the fact that \textit{ex vivo} scan is of chemically treated spinal cord at 7T while the \textit{in vivo} scans are of human brain at 3T. 

\begin{table*}[]
\centering
\textbf{Pearson Correlation of MWF Maps to Histology}
\vspace{2mm}

\centering
\vspace{2mm}
\begin{tabular}{lllll}
                & MIML            & NNLS-T          & NNLS-L          & GMF             \\
Tailored Bounds & (\textbf{0.54},5.63E-81) & (0.44,8.58E-53) & (0.45,5.51E-55) & (0.39,1.43E-39) \\
Fixed Bounds    & (\textbf{0.54},2.04E-81) & (0.44,8.58E-53) & (0.49,1.13E-64) & (0.39,1.43E-39)
\end{tabular}
\vspace{2mm}
\caption{Here we show a table of the Pearson correlations (with p-values) between the MWF maps and the histology map in a white matter mask. We note that the histology map is not a map of the myelin water fraction, but a map of the fraction of pixels in the histology which correspond to the myelin.  In either case of fixed or tailored bounds, the MWF map from MIML has the highest correlation to the histology. }
\label{table:histology_correlations}
\end{table*}

\subsubsection{Healthy Subjects}
\begin{figure*}
		\includegraphics[width=\textwidth]{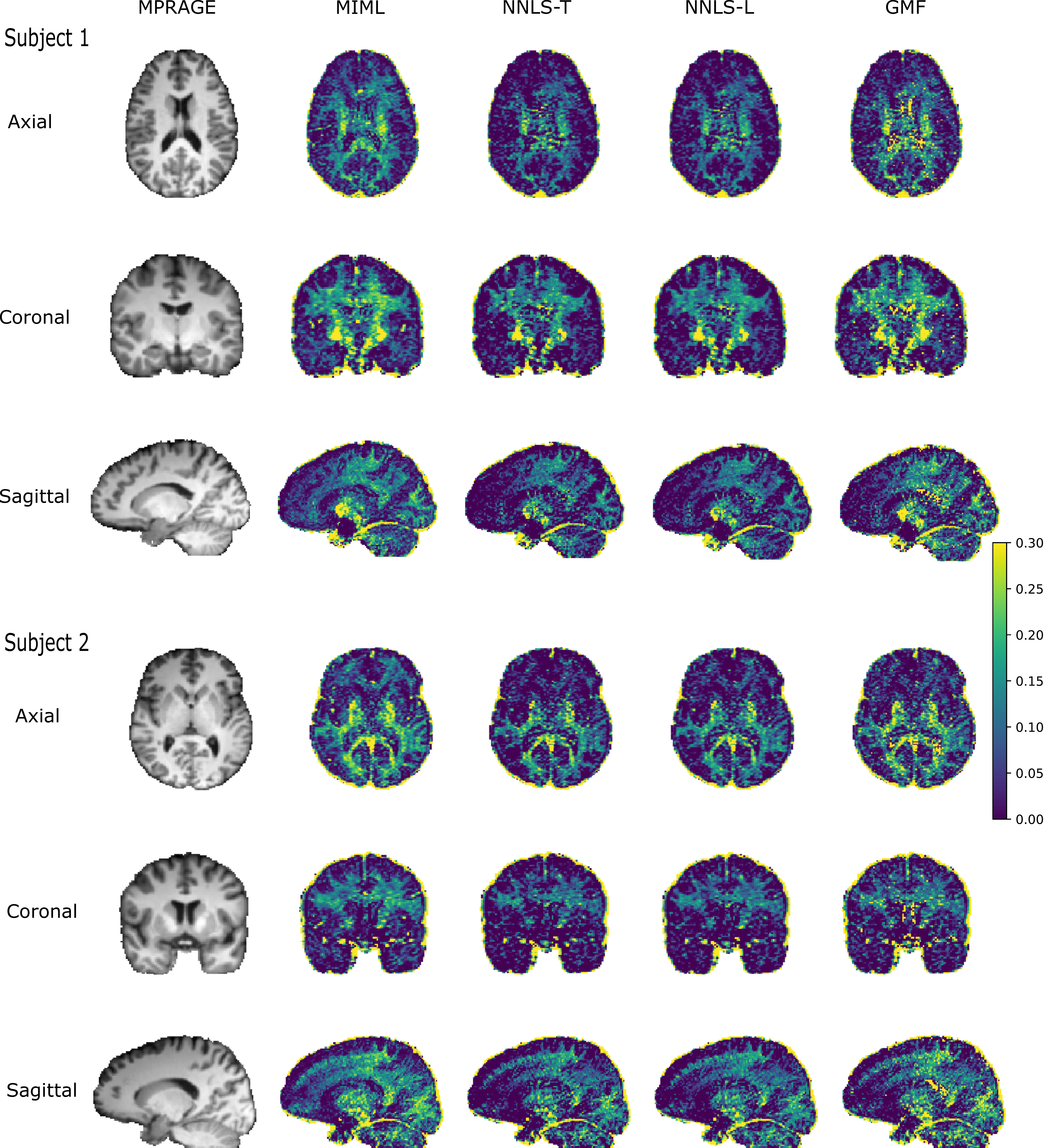}
		\caption{We show the MWF maps produced from each method, in the axial, coronal, and sagittal planes of two healthy subjects. On the left, we show the corresponding MPRAGE slice. Compared to the MPRAGE (where WM is hyper-intense), we can see that MIML most accurately and smoothly reproduces the extent of white matter, which is consistent with WM having relatively high MWF values. Particularly, the NNLS methods struggle in MWF recovery in the frontal part of the brain. GMF produces comparable to better MWF recovery than the NNLS methods, but with a noisier map. In addition, MIML has the least distortion in the ventricles.
		}
		\label{fig:healthy_subject_all_mwf}
\end{figure*}

\begin{figure*}
		\includegraphics[width=\textwidth]{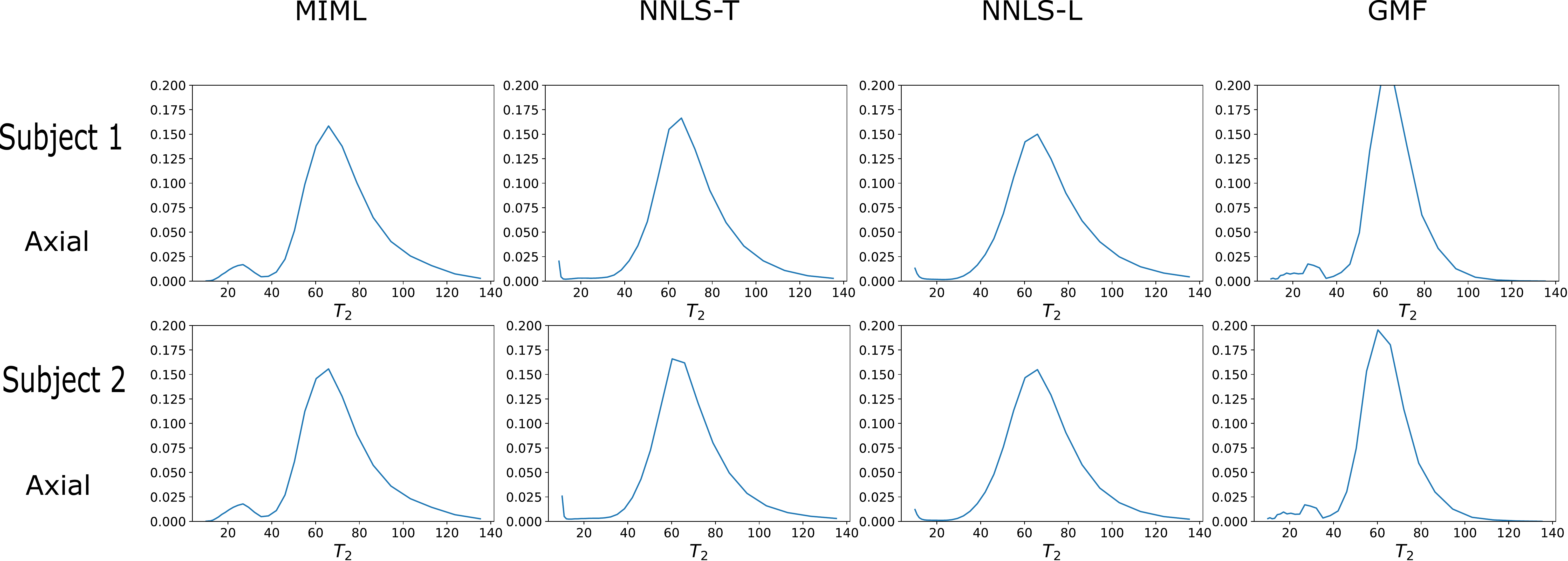}
		\caption{We show the mean distributions in the WM voxels of the axial slices of two healthy subjects for each method. We note that only MIML produces WM distributions with two distinct, well-separated lobes, and the myelin water peak in line with expectations at 3T. The NNLS methods and GMF recover the IE lobe well, but the myelin lobe is either irregular or appears at an extremely low $T_2$. 
		}
		\label{fig:healthy_subject_dist_axial}
\end{figure*}

In Fig. \ref{fig:healthy_subject_all_mwf} we show the MWF maps in axial, coronal, and sagittal slices for two healthy subjects with corresponding, registered MPRAGE images for comparison. In MPRAGE images, WM is hyperintense; hence, we treat the MPRAGE as a very rough proxy for the MWF map since MWF values are highest in the WM. Although the MWF maps are fairly similar, we see that MWF map of MIML most accurately and smoothly conforms to the MPRAGE image. The NNLS methods exhibit higher distortions, e.g. in the ventricles of subject 1, and difficulty in recovering the MWF in the frontal region of the brain. GMF produces maps comparable to the NNLS methods, albeit, looking noisier. 
We note that all methods exhibit lower MWFs in the frontal part of the brain as compared to other regions, which may stem from effects due to the gradient echo acquisition \citep{alonso2017field}.
In Fig. \ref{fig:healthy_subject_dist_axial}, we show the mean distributions over the WM voxels in the axial slices; for better visualization, we truncate the plots to a range encompassing WM $T_2$s (10-150ms). Only MIML produces WM distributions with two distinct, well-separated lobes corresponding to myelin water and the IES water as is expected from previous studies. Further, the peaks of the myelin water lobe and the IES water lobe correspond to the range expected at 3T. The NNLS methods recover the IES water lobe in line with expectations, but over-smooth the distribution in the region corresponding to myelin water, as was seen in the results on the synthetic data, with an implausible myelin water peak at 10ms. GMF also recovers the IE lobe in line with expectations, but produces a irregular lobe in the myelin region. 

In order to compare the MWF maps on regions of interest, and to conduct the scan-rescan analysis we did the following: in a first step, all the estimated MWF images for the 4 subjects were registered to the 'ICBM-DTI-81' white-matter tract labels atlas \citep{oishi2008human,mori2008stereotaxic} using the non-linear registration 'BSplineSyN' algorithm included in the ANTs software (https://github.com/ANTsX/ANTs). After visually inspecting the images, we removed small ROIs affected by registration errors and kept 44 tract labels showing a good anatomical agreement between the atlas and subject native spaces. Finally, the mean MWF value and the coefficient of variation of the MWF for each region of interest (ROI) was calculated for the scan and rescan maps from each method. A list of the ROIs can be found in Table 1 of the Supplementary material.
In Table \ref{table:coeff_of_var_roi}, we show the mean and standard deviation of the coefficient of variation of the MWF values in the ROIs. MIML produces significantly lower values for the mean and standard deviation of the coefficient of variation compared to those of the other methods, indicating that the MIML MWF map is smoother in the ROIs. The NNLS methods and GMF perform similarly. 

In Tables \ref{table:scan_rescan} and \ref{table:scan_rescan_pearson}, we show the results of our scan-rescan analysis over all four healthy subjects; we show a table of the mean and standard deviation of the absolute difference between the mean MWF values of the scan and rescan in the specified ROIs as well as a table of the Pearson correlation and linear regression coefficients between the mean MWF values of the scan and rescan in the specified ROIs. We can see that in general, GMF provides the smallest mean differences and highest Pearson correlations. In particular, it is difficult to rank MIML and the NNLS methods as they perform better/worse on different subjects. We note that GMF's superior reproducibility may stem from the lower flexibility in the fitting of the MWF, as compared to MIML and the NNLS methods. However, overall, the reproducibility of the methods is quite similar. 

\begin{table*}[]
\centering
\textbf{Mean and Standard Deviation of the Coefficient of Variation in WM ROIs}
\vspace{2mm}

\centering
\vspace{2mm}
\begin{tabular}{lllll}
          & MIML             & NNLS-T           & NNLS-L           & GMF              \\
Subject 1 & (\textbf{0.4},\textbf{0.12})  & (0.56,0.21) & (0.51,0.19) & (0.51,0.17)  \\
Subject 2 & (\textbf{0.44},\textbf{0.15}) & (0.58,0.24) & (0.55,0.23) & (0.53,0.22)  \\
Subject 3 & (\textbf{0.42},\textbf{0.15}) & (0.5,0.18)  & (0.48,0.18)   & (0.48,0.16) \\
Subject 4 & (\textbf{0.41},\textbf{0.14})  & (0.55,0.24) & (0.51,0.21) & (0.5,0.2)
\end{tabular}
\vspace{2mm}
\caption{Here we show a table of the mean and standard deviation of the coefficient of variation of the MWF values in the specified WM ROIs of the healthy subjects. Overall, MIML produces the smallest mean and standard deviation of the coefficient of variation in all subjects, indicating that the MWF map is smoother in the ROIs. The NNLS methods and GMF have comparable statistics for the coefficient of variation.  }
\label{table:coeff_of_var_roi}
\end{table*}

\begin{table*}[]
\centering
\textbf{Mean and Standard Deviation of MWF Differences between Scan and Rescan WM ROIs}
\vspace{2mm}

\centering
\vspace{2mm}
\begin{tabular}{lllll}
          & MIML             & NNLS-T           & NNLS-L           & GMF              \\
Subject 1 & (0.0067,0.0388)  & (0.0093,0.0353) & (0.0094,\textbf{0.0305}) & (\textbf{0.0055},0.03565)  \\
Subject 2 & (\textbf{0.0004},0.0448) & (0.0012,0.0473) & (0.001,0.04175) & (0.0044,\textbf{0.0454})  \\
Subject 3 & (0.0191,0.07255) & (0.0134,0.0755)  & (0.0176,0.0723)   & (\textbf{0.0108},\textbf{0.0712}) \\
Subject 4 & (0.0104,0.05725)  & (0.0107,0.05425) & (0.0103,\textbf{0.0504}) & (\textbf{0.0065},0.0561)  
\end{tabular}
\vspace{2mm}
\caption{Here we show a table of the mean and standard deviation of the absolute difference between the mean MWF values of the scan and rescan in specified ROIs for the healthy subjects. Overall, GMF has the smallest mean differences for 3/4 subjects with standard deviations comparable to those of other methods. The performance of MIML and the NNLS methods are overall quite similar; while MIML has the smallest mean difference on Subject 2, NNLS methods have smaller mean differences in Subjects 3 and 4. }
\label{table:scan_rescan}
\end{table*}

\begin{table*}[]
\centering
\textbf{Pearson Correlation and Linear Regression Coefficients between Scan and Rescan ROIs}
\vspace{2mm}

\centering
\vspace{2mm}
\begin{tabular}{lllll}
          & MIML                & NNLS-T             & NNLS-L              & GMF                  \\
Subject 1 & (0.92, 0.895,0.0069) & (0.91,0.887,0.0016) & (\textbf{0.93},0.901,0.00039) & (\textbf{0.93},0.85,0.013) \\
Subject 2 & (0.9,1.0039,0.00016)   & (0.87,0.98,0.00084)  & (0.89,1.022,-0.0034)  & (\textbf{0.91},0.99,0.00454)  \\
Subject 3 & (\textbf{0.77},0.74,0.0546)   & (0.7,0.649,0.0527)   & (0.72,0.685,0.052)   & (\textbf{0.77},0.76,0.0433)    \\
Subject 4 & (0.87,0.87,0.0087)   & (0.87,0.965,-0.0064)  & (\textbf{0.89},0.97,-0.0076)  & (\textbf{0.89},0.955,0.00014)   
\end{tabular}
\vspace{2mm}
\caption{Here we show a table of the Pearson correlation and the linear regression coefficients (slope and intercept) between the mean MWF values of the scan and rescan in specified ROIs for the healthy subjects. All Pearson correlations have $p$ values less than 0.01. Overall, all methods perform quite similarly. However, GMF has the best correlations between scans (by a small margin), with MIML and the NNLS methods performing similarly. }
\label{table:scan_rescan_pearson}
\end{table*}

\subsubsection{MS Subject}
\begin{figure*}
		\includegraphics[width=\textwidth]{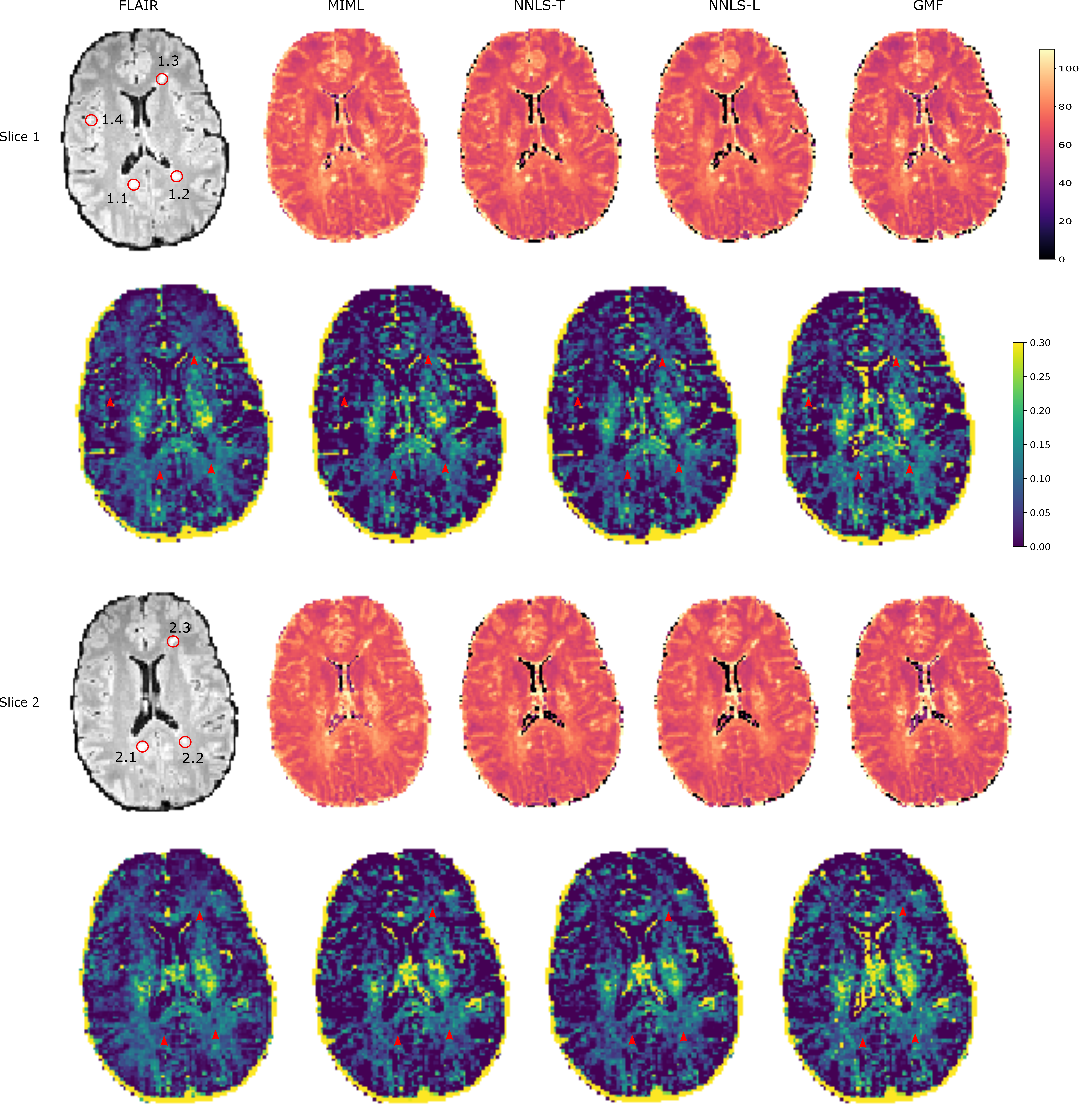}
		\caption{We show maps of the mean $T_2$ in each voxel within the range corresponding to the IE space (50-200ms) (1st and 3rd row), and MWF maps (2nd and 4th row) for two consectutive axial slices in the subject with MS. In addition, we show the corresponding slices on the FLAIR image with the MS lesions marked in red and labeled numerically. Regarding the mean $T_2$ maps, we can see that the all lesions but Lesion 2.2 can be seen as hyperintensities, with the maps very similar across all methods. Regarding the MWF maps, as in the healthy subjects, MIML most smoothly and accurately reconstructs the WM, with the other methods exhibiting more noisy maps with missing patches. In contrast to the mean $T_2$ maps, we can see that, for all methods, only lesion 1.2 can be distinctly seen from the MWF maps in Slice 1. In Slice 2, MIML provides the best lesion visualization due to better contrast between normal appearing tissue and lesions and a more smooth MWF map; in particular, lesions can clearly be delineated from close, adjacent structures in contrast to the NNLs methods (see Lesion 2.1,2.3).  See Fig. \ref{fig:ms_subjects_slice_2_mwf_zoom} for a closer look/analysis of the MWF maps compared to the lesions.
		}
		\label{fig:ms_subjects_mean_T2_MWF}
\end{figure*}

\begin{figure*}
		\includegraphics[width=15cm,height=21cm]{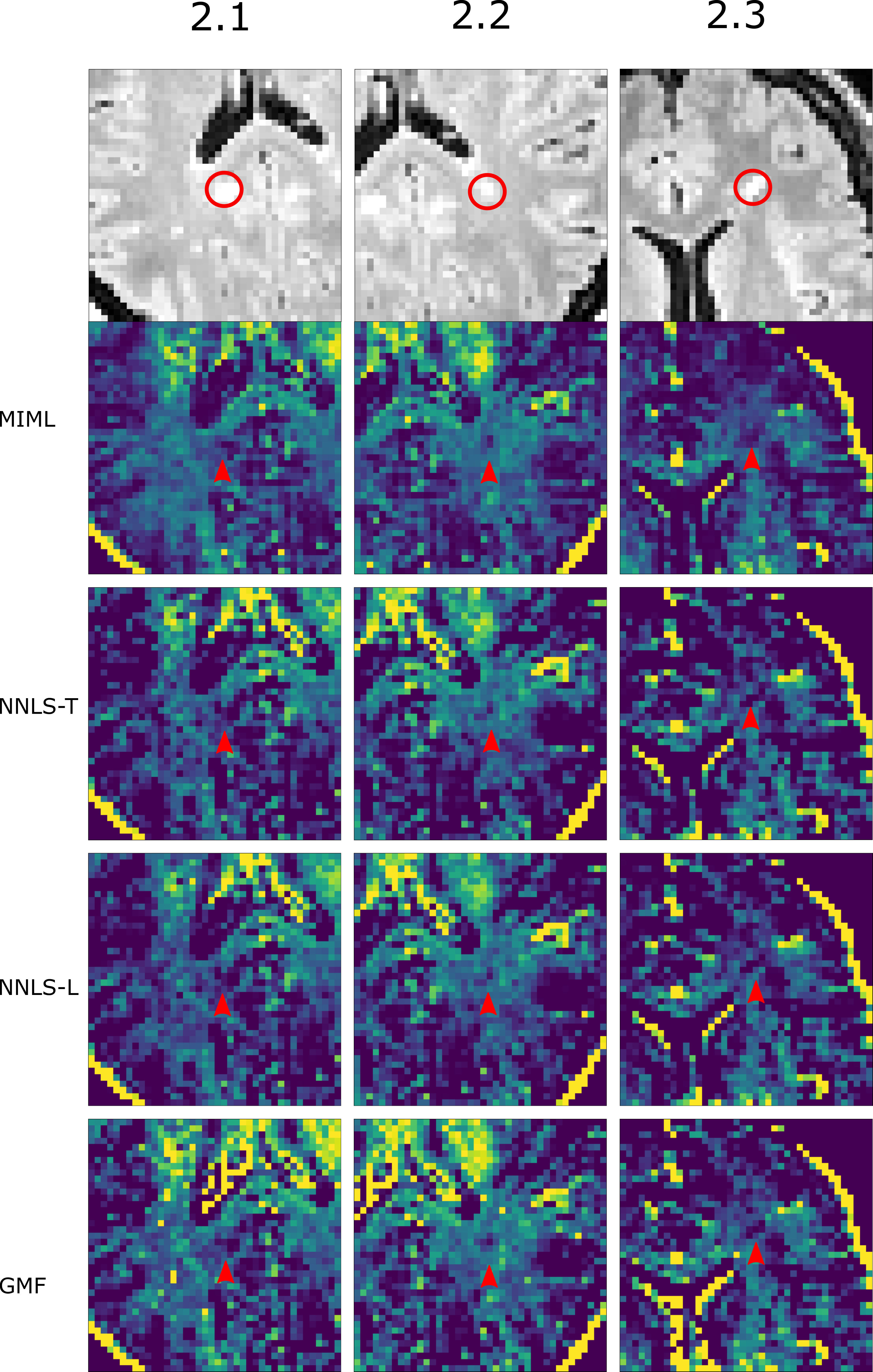}
		\caption{Here we zoom in on the lesions in slice 2 as well as the corresponding patches in the MWF map. We are able to see distinctly see all the lesions using the MIML MWF map; in particular, lesions 2.1 and 2.3 can be clearly distinguished from close, adjacent structures. Due to the lower contrast between normal appearing tissue and lesion tissue and noisier appearance in comparison to the MIML MWF map, lesions 2.1 and 2.2 are somewhat ambiguous on the NNLS maps; in particular, it appears the lesion 2.1 is exaggerated in size and mixed with the structure next to it. Similarly Lesion 2.3 is mixed with the structure next to it with the NNLS maps. The GMF MWF maps are similar to those of MIML, albeit noisier.
		}
		\label{fig:ms_subjects_slice_2_mwf_zoom}
\end{figure*}

 In Fig. \ref{fig:ms_subjects_mean_T2_MWF}, we show the maps of the geometric mean $T_2$ in the IE range of 50-200ms as well as the MWF maps of two consecutive, axial slices in a single subject. In addition, in Fig. \ref{fig:ms_subjects_slice_2_mwf_zoom}, we zoom in on the lesions in slice 2 for better visualization. We can see that for the mean $T_2$ maps, in all methods, almost all of the lesions in both slices can be clearly seen as hyperintensities  i.e. with increased mean IE $T_2$. Further, we see the maps are similar across the methods, with the main differences residing in the ventricles. Visualizing the lesions is far more difficult with MWF maps than with the mean $T_2$ maps, as the MWF maps are much noisier independent of the applied method. However, as with the healthy subjects, the MIML MWF map in both slices most smoothly and accurately conforms to the WM and the cortices, with the other methods exhibiting more variability and missing patches in the WM and worse delineation of the cortices; this occurs particularly in the frontal region. In Slice 1, in all methods, only Lesion 1.2 can be seen unambiguously. All methods show a dark spot near where Lesion 1.3 is expected, but not at the correct location; this may stem from small registration errors between the FLAIR and the multi-echo $T_2$ space. In Slice 2, all three lesions can be seen on the MIML MWF map with minimal ambiguity; in particular, in lesions 2.1 and 2.3, we can clearly delineate the lesions from very close, adjacent structures. Concerning the NNLS methods, it appears that Lesion 2.1 is exaggerated in size and mixed with the adjacent structure, making it difficult to delineate the lesion as the dark region is extended far beyond the lesion region on the FLAIR image. In addition, due to poor contrast between the normal-appearing tissue and lesion tissue/noise, it is difficult to identify Lesion 2.2 unambiguously with the NNLS methods. As with Lesion 2.1, Lesion 2.3 can be seen but is connected to the adjacent grey matter, making localization problematic. Further, we can see that the MWF in the lesion is comparable to the MWF of the normal-appearing, contralateral brain region, due to the poor MWF reconstruction. The GMF MWF map resembles the MIML MWF map albeit noisier/ with greater variability, making identification of the lesions more difficult.

\begin{figure*}
		\includegraphics[width=\textwidth,height=9.5cm]{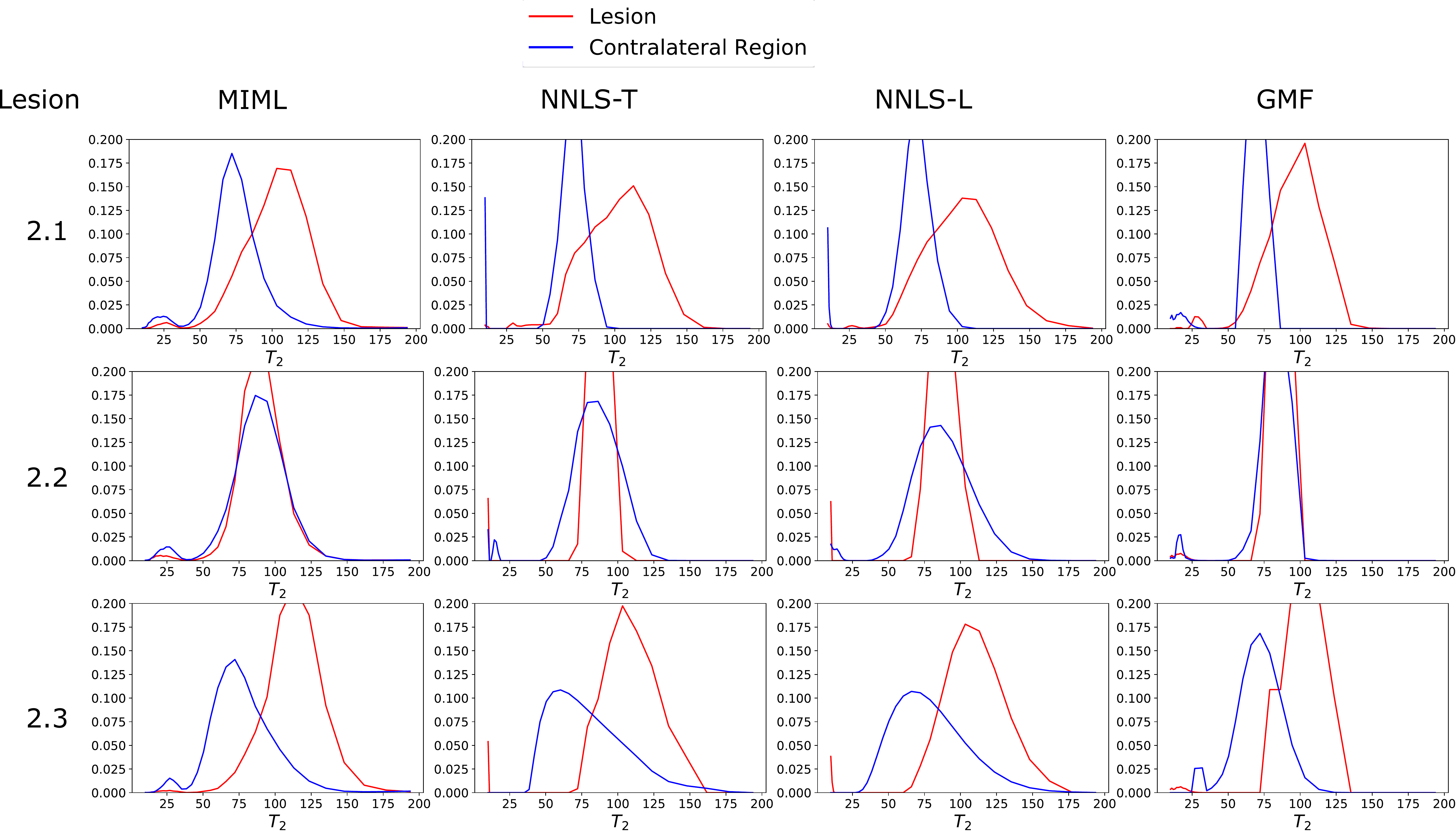}
		\caption{Here we compare the mean $T_2$ distributions for slice 2 within the lesion mask and the same mask translated to the normal appearing region contralateral to the lesion for each lesion and method. We can see that MIML consistently produces two distinct, well-separated lobes, and the myelin water peak in line with expectations at 3T. In general, the NNLS methods and GMF recover the IE lobe well, with occasional noise, but the myelin water lobes are over-smoothed with peaks at implausibly low values or irregular. We note that MIML finds a diminished myelin water lobe in the lesion as compared to the normal appearing tissue, in line with expectations of MS as a demyelinating disorder; in contrast, the NNLS methods in Lesions 2.2/2.3 exhibit larger myelin water lobes in lesion tissue as compared to normal appearing tissue. GMF performs similarly to MIML in this regard, albeit with more irregular distributions. 
		}
		\label{fig:ms_subjects_mean_dist_slice_2}
\end{figure*}

In addition to the mean $T_2$ and MWF maps, in Fig \ref{fig:ms_subjects_mean_dist_slice_2}, we compare, for Slice 2, the mean $T_2$ distributions in the lesion masks to the $T_2$ distributions in the normal appearing, contralateral regions. As in the healthy subjects, we can see that MIML consistently produces two distinct, well-separated lobes corresponding to myelin water and the IES water as is expected from previous studies. Further, the peaks of the myelin water lobe and the IES water lobe correspond to the range expected at 3T. The NNLS methods produce over-smoothed myelin water lobes with peaks occurring at implausibly low $T_2$ values. The IES water lobes are generally plausible, albeit with increased noise. GMF produces more plausible myelin water lobes than those of the NNLS methods, but the lobes are largely irregular.  MIML reconstructs a diminished myelin water lobe in the lesions as compared to the normal-appearing tissue, reflecting lower MWF; this is in line with expectations of MS as a demyelinating disorder. In contrast, the NNLS methods in Lesions 2.2/2.3 exhibit larger myelin water lobes in lesion tissue as compared to normal-appearing tissue, indicative of the poor MWF reconstruction in the normal-appearing tissue. GMF performs similarly to MIML with regard to the myelin water lobes, albeit with more irregular distributions. 

In conclusion, all methods perform similarly in detecting lesions from the mean $T_2$. However, MIML improves upon the NNLS methods and GMF in detecting lesions from MWF maps, by providing better contrast between lesions and normal appearing tissues, clearer delineation of lesions from adjacent structures, and smoother, more plausible reconstructions overall in the WM. From Fig. \ref{fig:ms_subjects_mean_dist_slice_2}, we confirm that the distributions of MIML are also more plausible and match the model upon which MIML was trained; in particular, the comparison of myelin water lobes of lesion and normal appearing tissue from MIML is consistent with the demyelinating nature of MS in contrast to that from the NNLS methods. Therefore, the performance of MIML meets or exceeds the performance of the other methods when used on a pathological case. 

From our results on real data, we see that MIML generalizes to different machines, different magnetic field strengths, and different sequences since it is trained on a model of the signal decay which is agnostic to these differences; MIML's performance on the real data shows its potential for multi-component $T_2$ relaxometry at clinically achievable SNRs in high resolution scans.   

\subsection{Computation Time}
Here we provide a brief overview of the computational cost of the different methods. For consistent comparison, we used one computer using Ubuntu 18.04 with an Intel Xeon CPU E5-1650v4 running at 3.6 GHz with 12 available threads to run parallelized whole-brain reconstructions on four of the healthy subjects (matrix size 144x126x84) using MIML, NNLS-T, NNLS-L, and GMF; we recorded the time to completion and show the average computation time for each method in Table \ref{table:computation_time}. We can see that MIML is 1 to 4 orders of magnitude faster than the other methods.

\begin{table}[]
\centering
\textbf{Average Computation for Whole Brain}
\vspace{2mm}

\centering
\vspace{2mm}
\begin{tabular}{lllll}
                         & MIML & NNLS-T & NNLS-L & GMF       \\
Time & 34s  & 752s   & 701.2s & 159382.4s
\end{tabular}
\vspace{2mm}
\caption{Here we show a table of the average computation time for whole brain reconstructions on the healthy subjects using a single computer. We see that MIML is orders of magnitude faster than the other methods. }
\label{table:computation_time}
\end{table}

\section{Discussion}
Overall, from our evaluation on synthetic data, an \textit{ex vivo} scan and \textit{in vivo} scans (healthy and pathological), we conclude that MIML provides fast, noise-robust, and plausible reconstructions of $T_2$ distributions, with potential for use in myelin water fraction mapping. We attribute the performance of our method to the blending of the advantages of machine learning, parametric, and non-parametric methods. We note that our approach is essentially using machine learning to solve the inverse problem of parametric approaches, albeit expressing the solution non-parametrically. We view our approach as an extension of the recent progress in using machine learning to solve inverse problems in many domains \citep{adler2017solving}.
By using machine learning, our method is much faster than standard parametric or non-parametric approaches. By training on solely simulated data, our approach does not require expensive, \textit{in vivo} acquisitions for training data, nor the need for multiple scans to adapt to different machines or sequences. Further, this allows for noise-robust reconstruction by training the network on simulated signals with an SNR range and noise model corresponding to those from clinical scans. By generating the simulations guided by biophysical models, we can simultaneously retain stability in the reconstruction by constraining the space of $T_2$ distributions while not being restricted to a specific number of water pools at inference time. Further, the produced distributions are implicitly constrained to have a plausible, lobular structure (as in parametric approaches), which makes the interpretation of parameters of interest such as the MWF consistent with past studies, in contrast to potential irregular distributions from non-parametric methods. The trained MIML model and code for generating the synthetic data and training the model will be available at the following website: \href{https://github.com/thomas-yu-epfl/Model\_Informed\_Machine\_Learning.}{https://github.com/thomas-yu-epfl/Model\_Informed\_Machine\_Learning.} 

However, our current approach has several limitations. First, while we attempted to be as comprehensive as possible in the simulated dataset, advances in biophysical modelling make it possible that there are additional relevant water pools to be estimated. Second, while we fixed the Rician noise model for the training signals, with a fixed SNR range of 80-200, we note that in some sequences, more complex noise models such as the non-central chi distribution \citep{aja2016statistical} with different SNR ranges may also be appropriate. Third, we only consider 32-echo sequences in this work. Fourth, we use a fixed, logarithmic $T_2$ discretization consisting of 60 points from 10ms to 2000ms for both our method and the NNLS methods. However, finer or coarser discretizations could also have been used. Finally, there may be relevant physical effects such as magnetization transfer \citep{sled2018modelling,malik2018extended} which, if modelled in the dataset, could improve the reconstructions. 
However, we highlight the flexibility and modularity of our approach for accounting for these limitations. Additional water pools can be easily added to the training dataset. The noise model and SNR range used in training can be swapped out for different noise models and SNRs. A sequence with a different number of echoes can be accommodated by reconstructing the dataset with the required number of echos and retraining the network. Different $T_2$ discretizations would simply require downsampling of the high resolution $T_2$ distributions in our dataset to match the new discretization, with subsequent retraining of the network. More advanced physical modelling can be added to the generation of new datasets. As the training of the network is quite fast (15 min on a laptop GPU), the bottleneck for addressing these limitations is the dataset generation (24 hours on 46 CPU threads). However, while we generated our dataset on CPU, GPU acceleration of the EPG formalism can potentially speedup dataset generation significantly \citep{wang2020snapmrf}. 

As for future work: in this paper, we did not study the impact of denoising the data on the reconstruction performance of the methods compared. This is first because in our overview of the literature, we found that presenting results on denoised data is not typical unless the subject of the paper is denoising. Second, the type of denoising, setting of denoising parameters, and accounting for potential biases due to denoising all require careful justification and study, which we felt was out of the scope of this paper, which introduces a proof of concept. However, we note that in the MS data, particularly for the NNLS methods, ostensibly normal appearing regions of the brain had unusually low MWF values, sometimes less than that predicted for the lesion. These areas of unusually low MWF values could also be seen in the scans of healthy subjects. These may be due to, in part, instability/ill-posedness in the estimation due to comparatively low SNRs in the \textit{in vivo} scans; the \textit{in vivo} scans we used have fairly high resolution (1.6-1.8mm) and are isotropic, while typical scans in the literature generally use much thicker slices ( $\geq$ 2mm) along the axial direction \citep{alonso2015mri,prasloski2012applications}. We note that both distributions and MWF maps from the NNLS methods were more plausible in the \textit{ex vivo} scan, where the SNR was much higher. This is consistent with the observations in \citep{wiggerman} concerning the noise dependence of NNLS methods. Future studies will be conducted to study the impact of denoising algorithms such as PCA denoising \citep{does2019evaluation}, or the NESMA filter \citep{bouhrara2018use} on MIML as well as other methods, and any effect this has on their comparison. 

Our method, as well as the other methods compared to in this work, reconstruct the $T_2$ distribution in each voxel separately. However, there are parametric and non-parametric approaches to $T_2$ relaxometry which use spatial regularization \citep{el2020multi,hwang2009improved,kumar2018using}. These approaches assume that voxels spatially close to each other should also have similar reconstructions; hence, they perform reconstructions on groups of adjacent voxels simultaneously, with constraints that limit the variation of the reconstructions over the group. In addition, another approach estimates over groups of voxels by assuming the joint sparsity of the distributions in a region of interest \citep{nagtegaal2020myelin}. In future work, we will study how regularization/simultaneous fitting over regions of interest can be incorporated into our machine-learning framework as well as its effects on distribution reconstruction. 

In this paper, we tested our method on two types of sequences: a multi-echo spin echo sequence and a 3D gradient and spin echo sequence. While in principle our approach is agnostic to the sequence used, in the future we will further validate our method on data from other sequences such as the $T_2$ prepared gradient echo sequences \citep{nguyen2012t2prep}.

We note that our approach is most similar to multi-component $T_2$ mapping with Magnetic Resonance Fingerprinting (MRF)  \citep{tang2018multicompartment,mcgivney2018bayesian}, where a pre-computed dictionary of signals is generated according to a pseudo-random sequence, and the voxel signal is matched to the $T_2,T_1,$ and proton density parameters from the dictionary which generates the most similar signal. However, these approaches estimate a single $T_2$ for each compartment. As with NNLS, these methods require relatively high SNR. Furthermore, at this time, we are not aware of any works which accelerate multi-component $T_2$ mapping with MRF through machine learning, though this has been done with standard MRF \citep{hoppe2017deep}. Finally, our approach uses spin-echo sequences targeted for solely $T_2$ estimation while MRF uses pseudo-random sequences to estimate multiple parameters simultaneously. In the future, it would be interesting to compare results from multi-component $T_2$ mapping with MRF and our approach. 

Finally, we note that using more advanced neural networks such as Long short term memory (LSTM) networks \citep{hochreiter1997long}, which are suitable for time series data, may offer improved reconstructions. 

\section{Conclusion}

In this work, we presented Model-Informed Machine Learning (MIML), an approach for estimating $T_2$ distributions from MRI signals using a neural network trained on synthetic data derived from biophysical models. Through our evaluations on synthetic data, an \textit{ex vivo} scan, as well as healthy and pathological \textit{in vivo} data, we show that MIML provides more robust, accurate, and plausible $T_2$ distributions than standard parametric and non-parametric methods across a wide range of SNRs. We show that MWF maps derived from MIML show the highest conformity to anatomical scans, have the greatest correlation to  a histological map of myelin volume, and improve upon the lesion visualization capabilities of other methods, with better contrast between lesions and normal-appearing tissue as well as clearer delineation between lesions and close adjacent structures. The code for generating the datasets and training the network will be made available at \href{https://github.com/thomas-yu-epfl/Model\_Informed\_Machine\_Learning.}{https://github.com/thomas-yu-epfl/Model\_Informed\_Machine\_Learning.}

\section*{Acknowledgments}
This project is supported by the European Union's Horizon 2020 research and innovation programme under the Marie Sklodowska-Curie project TRABIT (agreement No 765148 to TY) and by the Swiss National Science Foundation (SNSF, Ambizione grant PZ00P2\_185814/1 to EJC-R). It has received funding from the European Union’s Horizon 2020 research and innovation programme under the Marie Skłodowska-Curie grant agreement No 754462 (to MP) and the Strategic Focal Area “Personalized Healthcare and Related Technologies (PHRT)” of the ETH domain (grant 2018-425 to EFG), as well as the Centre for Biomedical Imaging (CIBM) of the University of Lausanne, the Swiss Federal Institute of Technology Lausanne, the Lausanne University Hospital (to MBC and J-PT), and the University of Geneva and Geneva University Hospital (to MBC). In addition, this project was supported by Swiss National Funds PZ00P3\_154508, PZ00P3\_131914 and PP00P3\_176984 (CG,MW,MB).

\bibliographystyle{model1-num-names.bst}\biboptions{authoryear}
\bibliography{refs}

\begin{thebibliography}{68}
\expandafter\ifx\csname natexlab\endcsname\relax\def\natexlab#1{#1}\fi
\providecommand{\bibinfo}[2]{#2}
\ifx\xfnm\relax \def\xfnm[#1]{\unskip,\space#1}\fi
\bibitem[{Haacke et~al.(1999)Haacke, Brown, Thompson, Venkatesan, Thomphson,
  Venkatesan, Haacke, Brown, and Thompson}]{haacke1999magnetic}
\bibinfo{author}{E.~M. Haacke}, \bibinfo{author}{R.~W. Brown},
  \bibinfo{author}{M.~R. Thompson}, \bibinfo{author}{R.~Venkatesan},
  \bibinfo{author}{M.~Thomphson}, \bibinfo{author}{M.~Venkatesan},
  \bibinfo{author}{M.~Haacke}, \bibinfo{author}{W.~Brown},
  \bibinfo{author}{M.~Thompson},
\newblock \bibinfo{title}{Magnetic resonance imaging: physical principles and
  sequence design}  (\bibinfo{year}{1999}).
\bibitem[{Menon and Allen(1991)}]{menon1991application}
\bibinfo{author}{R.~Menon}, \bibinfo{author}{P.~Allen},
\newblock \bibinfo{title}{Application of continuous relaxation time
  distributions to the fitting of data from model systmes and excised tissue},
\newblock \bibinfo{journal}{Magnetic resonance in medicine}
  \bibinfo{volume}{20} (\bibinfo{year}{1991}) \bibinfo{pages}{214--227}.
\bibitem[{Bloch(1946)}]{bloch1946nuclear}
\bibinfo{author}{F.~Bloch},
\newblock \bibinfo{title}{Nuclear induction},
\newblock \bibinfo{journal}{Physical review} \bibinfo{volume}{70}
  (\bibinfo{year}{1946}) \bibinfo{pages}{460}.
\bibitem[{Prasloski et~al.(2012)Prasloski, M{\"a}dler, Xiang, MacKay, and
  Jones}]{prasloski2012applications}
\bibinfo{author}{T.~Prasloski}, \bibinfo{author}{B.~M{\"a}dler},
  \bibinfo{author}{Q.-S. Xiang}, \bibinfo{author}{A.~MacKay},
  \bibinfo{author}{C.~Jones},
\newblock \bibinfo{title}{Applications of stimulated echo correction to
  multicomponent t2 analysis},
\newblock \bibinfo{journal}{Magnetic resonance in medicine}
  \bibinfo{volume}{67} (\bibinfo{year}{2012}) \bibinfo{pages}{1803--1814}.
\bibitem[{Hennig(1988)}]{hennig1988multiecho}
\bibinfo{author}{J.~Hennig},
\newblock \bibinfo{title}{Multiecho imaging sequences with low refocusing flip
  angles},
\newblock \bibinfo{journal}{Journal of Magnetic Resonance (1969)}
  \bibinfo{volume}{78} (\bibinfo{year}{1988}) \bibinfo{pages}{397--407}.
\bibitem[{Neumann et~al.(2014)Neumann, Blaimer, Jakob, and
  Breuer}]{neumann2014simple}
\bibinfo{author}{D.~Neumann}, \bibinfo{author}{M.~Blaimer},
  \bibinfo{author}{P.~M. Jakob}, \bibinfo{author}{F.~A. Breuer},
\newblock \bibinfo{title}{Simple recipe for accurate t 2 quantification with
  multi spin-echo acquisitions},
\newblock \bibinfo{journal}{Magnetic Resonance Materials in Physics, Biology
  and Medicine} \bibinfo{volume}{27} (\bibinfo{year}{2014})
  \bibinfo{pages}{567--577}.
\bibitem[{MacKay and Laule(2007)}]{mackay2007myelin}
\bibinfo{author}{A.~L. MacKay}, \bibinfo{author}{C.~Laule},
\newblock \bibinfo{title}{Myelin water imaging},
\newblock \bibinfo{journal}{eMagRes}  (\bibinfo{year}{2007}).
\bibitem[{Mackay et~al.(1994)Mackay, Whittall, Adler, Li, Paty, and
  Graeb}]{mackay1994vivo}
\bibinfo{author}{A.~Mackay}, \bibinfo{author}{K.~Whittall},
  \bibinfo{author}{J.~Adler}, \bibinfo{author}{D.~Li},
  \bibinfo{author}{D.~Paty}, \bibinfo{author}{D.~Graeb},
\newblock \bibinfo{title}{In vivo visualization of myelin water in brain by
  magnetic resonance},
\newblock \bibinfo{journal}{Magnetic resonance in medicine}
  \bibinfo{volume}{31} (\bibinfo{year}{1994}) \bibinfo{pages}{673--677}.
\bibitem[{Whittall et~al.(1997)Whittall, Mackay, Graeb, Nugent, Li, and
  Paty}]{whittall1997vivo}
\bibinfo{author}{K.~P. Whittall}, \bibinfo{author}{A.~L. Mackay},
  \bibinfo{author}{D.~A. Graeb}, \bibinfo{author}{R.~A. Nugent},
  \bibinfo{author}{D.~K. Li}, \bibinfo{author}{D.~W. Paty},
\newblock \bibinfo{title}{In vivo measurement of t2 distributions and water
  contents in normal human brain},
\newblock \bibinfo{journal}{Magnetic resonance in medicine}
  \bibinfo{volume}{37} (\bibinfo{year}{1997}) \bibinfo{pages}{34--43}.
\bibitem[{Vasilescu et~al.(1978)Vasilescu, Katona, Simplaceanu, and
  Demco}]{vasilescu1978water}
\bibinfo{author}{V.~Vasilescu}, \bibinfo{author}{E.~Katona},
  \bibinfo{author}{V.~Simplaceanu}, \bibinfo{author}{D.~Demco},
\newblock \bibinfo{title}{Water compartments in the myelinated nerve. iii.
  pulsed nmr result},
\newblock \bibinfo{journal}{Experientia} \bibinfo{volume}{34}
  (\bibinfo{year}{1978}) \bibinfo{pages}{1443--1444}.
\bibitem[{Menon et~al.(1992)Menon, Rusinko, and Allen}]{menon1992proton}
\bibinfo{author}{R.~Menon}, \bibinfo{author}{M.~Rusinko},
  \bibinfo{author}{P.~Allen},
\newblock \bibinfo{title}{Proton relaxation studies of water
  compartmentalization in a model neurological system},
\newblock \bibinfo{journal}{Magnetic resonance in medicine}
  \bibinfo{volume}{28} (\bibinfo{year}{1992}) \bibinfo{pages}{264--274}.
\bibitem[{Raj et~al.(2014)Raj, Pandya, Shen, LoCastro, Nguyen, and
  Gauthier}]{raj2014multi}
\bibinfo{author}{A.~Raj}, \bibinfo{author}{S.~Pandya},
  \bibinfo{author}{X.~Shen}, \bibinfo{author}{E.~LoCastro},
  \bibinfo{author}{T.~D. Nguyen}, \bibinfo{author}{S.~A. Gauthier},
\newblock \bibinfo{title}{Multi-compartment t2 relaxometry using a spatially
  constrained multi-gaussian model},
\newblock \bibinfo{journal}{PLoS One} \bibinfo{volume}{9}
  (\bibinfo{year}{2014}).
\bibitem[{Du et~al.(2007)Du, Chu, Hwang, Brown, Kleinschmidt-DeMasters, Singel,
  and Simon}]{du2007fast}
\bibinfo{author}{Y.~P. Du}, \bibinfo{author}{R.~Chu},
  \bibinfo{author}{D.~Hwang}, \bibinfo{author}{M.~S. Brown},
  \bibinfo{author}{B.~K. Kleinschmidt-DeMasters}, \bibinfo{author}{D.~Singel},
  \bibinfo{author}{J.~H. Simon},
\newblock \bibinfo{title}{Fast multislice mapping of the myelin water fraction
  using multicompartment analysis of t decay at 3t: A preliminary postmortem
  study},
\newblock \bibinfo{journal}{Magnetic Resonance in Medicine: An Official Journal
  of the International Society for Magnetic Resonance in Medicine}
  \bibinfo{volume}{58} (\bibinfo{year}{2007}) \bibinfo{pages}{865--870}.
\bibitem[{Yu et~al.(2019)Yu, Pizzolato, Canales-Rodr{\'\i}guez, and
  Thiran}]{yu2019robust}
\bibinfo{author}{T.~Yu}, \bibinfo{author}{M.~Pizzolato}, \bibinfo{author}{E.~J.
  Canales-Rodr{\'\i}guez}, \bibinfo{author}{J.-P. Thiran},
\newblock \bibinfo{title}{Robust t 2 relaxometry with hamiltonian mcmc for
  myelin water fraction estimation},
\newblock in: \bibinfo{booktitle}{2019 IEEE 16th International Symposium on
  Biomedical Imaging (ISBI 2019)}, \bibinfo{organization}{IEEE}, pp.
  \bibinfo{pages}{1813--1817}.
\bibitem[{Chatterjee et~al.(2018)Chatterjee, Commowick, Afacan, Warfield, and
  Barillot}]{chatterjee2018multi}
\bibinfo{author}{S.~Chatterjee}, \bibinfo{author}{O.~Commowick},
  \bibinfo{author}{O.~Afacan}, \bibinfo{author}{S.~K. Warfield},
  \bibinfo{author}{C.~Barillot},
\newblock \bibinfo{title}{Multi-compartment model of brain tissues from t2
  relaxometry mri using gamma distribution},
\newblock in: \bibinfo{booktitle}{2018 IEEE 15th International Symposium on
  Biomedical Imaging (ISBI 2018)}, \bibinfo{organization}{IEEE}, pp.
  \bibinfo{pages}{141--144}.
\bibitem[{Akhondi-Asl et~al.(2014)Akhondi-Asl, Afacan, Mulkern, and
  Warfield}]{akhondi2014t}
\bibinfo{author}{A.~Akhondi-Asl}, \bibinfo{author}{O.~Afacan},
  \bibinfo{author}{R.~V. Mulkern}, \bibinfo{author}{S.~K. Warfield},
\newblock \bibinfo{title}{T 2-relaxometry for myelin water fraction extraction
  using wald distribution and extended phase graph},
\newblock in: \bibinfo{booktitle}{International Conference on Medical Image
  Computing and Computer-Assisted Intervention},
  \bibinfo{organization}{Springer}, pp. \bibinfo{pages}{145--152}.
\bibitem[{Bj{\"o}rk et~al.(2016)Bj{\"o}rk, Zachariah, Kullberg, and
  Stoica}]{bjork2016multicomponent}
\bibinfo{author}{M.~Bj{\"o}rk}, \bibinfo{author}{D.~Zachariah},
  \bibinfo{author}{J.~Kullberg}, \bibinfo{author}{P.~Stoica},
\newblock \bibinfo{title}{A multicomponent t2 relaxometry algorithm for myelin
  water imaging of the brain},
\newblock \bibinfo{journal}{Magnetic resonance in medicine}
  \bibinfo{volume}{75} (\bibinfo{year}{2016}) \bibinfo{pages}{390--402}.
\bibitem[{Prange and Song(2009)}]{prange2009quantifying}
\bibinfo{author}{M.~Prange}, \bibinfo{author}{Y.-Q. Song},
\newblock \bibinfo{title}{Quantifying uncertainty in nmr t2 spectra using monte
  carlo inversion},
\newblock \bibinfo{journal}{Journal of Magnetic Resonance}
  \bibinfo{volume}{196} (\bibinfo{year}{2009}) \bibinfo{pages}{54--60}.
\bibitem[{Alonso-Ortiz et~al.(2015)Alonso-Ortiz, Levesque, and
  Pike}]{alonso2015mri}
\bibinfo{author}{E.~Alonso-Ortiz}, \bibinfo{author}{I.~R. Levesque},
  \bibinfo{author}{G.~B. Pike},
\newblock \bibinfo{title}{Mri-based myelin water imaging: a technical review},
\newblock \bibinfo{journal}{Magnetic resonance in medicine}
  \bibinfo{volume}{73} (\bibinfo{year}{2015}) \bibinfo{pages}{70--81}.
\bibitem[{Lawson and Hanson(1995)}]{lawson1995solving}
\bibinfo{author}{C.~L. Lawson}, \bibinfo{author}{R.~J. Hanson},
  \bibinfo{title}{Solving least squares problems}, volume~\bibinfo{volume}{15},
  \bibinfo{publisher}{Siam}, \bibinfo{year}{1995}.
\bibitem[{Kroeker and Henkelman(1986)}]{kroeker1986analysis}
\bibinfo{author}{R.~M. Kroeker}, \bibinfo{author}{R.~M. Henkelman},
\newblock \bibinfo{title}{Analysis of biological nmr relaxation data with
  continuous distributions of relaxation times},
\newblock \bibinfo{journal}{Journal of Magnetic Resonance (1969)}
  \bibinfo{volume}{69} (\bibinfo{year}{1986}) \bibinfo{pages}{218--235}.
\bibitem[{Laule et~al.(2006)Laule, Leung, Li, Traboulsee, Paty, MacKay, and
  Moore}]{laule2006myelin}
\bibinfo{author}{C.~Laule}, \bibinfo{author}{E.~Leung}, \bibinfo{author}{D.~K.
  Li}, \bibinfo{author}{A.~Traboulsee}, \bibinfo{author}{D.~Paty},
  \bibinfo{author}{A.~MacKay}, \bibinfo{author}{G.~R. Moore},
\newblock \bibinfo{title}{Myelin water imaging in multiple sclerosis:
  quantitative correlations with histopathology},
\newblock \bibinfo{journal}{Multiple Sclerosis Journal} \bibinfo{volume}{12}
  (\bibinfo{year}{2006}) \bibinfo{pages}{747--753}.
\bibitem[{Graham et~al.(1996)Graham, Stanchev, and Bronskill}]{Graham1996}
\bibinfo{author}{S.~J. Graham}, \bibinfo{author}{P.~L. Stanchev},
  \bibinfo{author}{M.~J. Bronskill},
\newblock \bibinfo{title}{Criteria for analysis of multicomponent tissue t2
  relaxation data},
\newblock \bibinfo{journal}{Magnetic Resonance in Medicine}
  \bibinfo{volume}{35} (\bibinfo{year}{1996}) \bibinfo{pages}{370--378}.
\bibitem[{Andrews et~al.(2005)Andrews, Lancaster, Dodd, Contreras-Sesvold, and
  Fox}]{Andrews2005pools}
\bibinfo{author}{T.~Andrews}, \bibinfo{author}{J.~L. Lancaster},
  \bibinfo{author}{S.~J. Dodd}, \bibinfo{author}{C.~Contreras-Sesvold},
  \bibinfo{author}{P.~T. Fox},
\newblock \bibinfo{title}{Testing the three-pool white matter model adapted for
  use with t2 relaxometry},
\newblock \bibinfo{journal}{Magnetic Resonance in Medicine: An Official Journal
  of the International Society for Magnetic Resonance in Medicine}
  \bibinfo{volume}{54} (\bibinfo{year}{2005}) \bibinfo{pages}{449--454}.
\bibitem[{Wiggermann et~al.(2020)Wiggermann, Vavasour, Kolind, MacKay, Helms,
  and Rauscher}]{wiggerman}
\bibinfo{author}{V.~Wiggermann}, \bibinfo{author}{I.~M. Vavasour},
  \bibinfo{author}{S.~Kolind}, \bibinfo{author}{A.~L. MacKay},
  \bibinfo{author}{G.~Helms}, \bibinfo{author}{A.~Rauscher},
\newblock \bibinfo{title}{Non-negative least squares computation for in vivo
  myelin mapping using simulated multi-echo spin-echo t2 decay data},
\newblock \bibinfo{journal}{NMR in Biomedicine}  (\bibinfo{year}{2020})
  \bibinfo{pages}{e4277}.
\bibitem[{Kumar et~al.(2012)Kumar, Nguyen, Gauthier, and Raj}]{Kumar2012bayes}
\bibinfo{author}{D.~Kumar}, \bibinfo{author}{T.~D. Nguyen},
  \bibinfo{author}{S.~A. Gauthier}, \bibinfo{author}{A.~Raj},
\newblock \bibinfo{title}{Bayesian algorithm using spatial priors for
  multiexponential t2 relaxometry from multiecho spin echo mri},
\newblock \bibinfo{journal}{Magnetic resonance in medicine}
  \bibinfo{volume}{68} (\bibinfo{year}{2012}) \bibinfo{pages}{1536--1543}.
\bibitem[{Raj et~al.(2014)Raj, Pandya, Shen, LoCastro, Nguyen, and
  Gauthier}]{Raj2014spatial}
\bibinfo{author}{A.~Raj}, \bibinfo{author}{S.~Pandya},
  \bibinfo{author}{X.~Shen}, \bibinfo{author}{E.~LoCastro},
  \bibinfo{author}{T.~D. Nguyen}, \bibinfo{author}{S.~A. Gauthier},
\newblock \bibinfo{title}{Multi-compartment t2 relaxometry using a spatially
  constrained multi-gaussian model},
\newblock \bibinfo{journal}{PLoS One} \bibinfo{volume}{9}
  (\bibinfo{year}{2014}) \bibinfo{pages}{e98391}.
\bibitem[{Lee et~al.(2019)Lee, Lee, Choi, Shin, Shin, and
  Lee}]{lee2019artificial}
\bibinfo{author}{J.~Lee}, \bibinfo{author}{D.~Lee}, \bibinfo{author}{J.~Y.
  Choi}, \bibinfo{author}{D.~Shin}, \bibinfo{author}{H.-G. Shin},
  \bibinfo{author}{J.~Lee},
\newblock \bibinfo{title}{Artificial neural network for myelin water imaging},
\newblock \bibinfo{journal}{Magnetic resonance in medicine}
  (\bibinfo{year}{2019}).
\bibitem[{Liu et~al.(2020)Liu, Xiang, Tam, Dvorak, MacKay, Kolind, Traboulsee,
  Vavasour, Li, Kramer, and Laule}]{LIU2020116551}
\bibinfo{author}{H.~Liu}, \bibinfo{author}{Q.-S. Xiang},
  \bibinfo{author}{R.~Tam}, \bibinfo{author}{A.~V. Dvorak},
  \bibinfo{author}{A.~L. MacKay}, \bibinfo{author}{S.~H. Kolind},
  \bibinfo{author}{A.~Traboulsee}, \bibinfo{author}{I.~M. Vavasour},
  \bibinfo{author}{D.~K. Li}, \bibinfo{author}{J.~K. Kramer},
  \bibinfo{author}{C.~Laule},
\newblock \bibinfo{title}{Myelin water imaging data analysis in less than one
  minute},
\newblock \bibinfo{journal}{NeuroImage} \bibinfo{volume}{210}
  (\bibinfo{year}{2020}) \bibinfo{pages}{116551}.
\bibitem[{Prasloski et~al.(2012)Prasloski, Rauscher, MacKay, Hodgson, Vavasour,
  Laule, and M{\"a}dler}]{prasloski2012rapid}
\bibinfo{author}{T.~Prasloski}, \bibinfo{author}{A.~Rauscher},
  \bibinfo{author}{A.~L. MacKay}, \bibinfo{author}{M.~Hodgson},
  \bibinfo{author}{I.~M. Vavasour}, \bibinfo{author}{C.~Laule},
  \bibinfo{author}{B.~M{\"a}dler},
\newblock \bibinfo{title}{Rapid whole cerebrum myelin water imaging using a 3d
  grase sequence},
\newblock \bibinfo{journal}{Neuroimage} \bibinfo{volume}{63}
  (\bibinfo{year}{2012}) \bibinfo{pages}{533--539}.
\bibitem[{Rosenblatt(1958)}]{rosenblatt1958perceptron}
\bibinfo{author}{F.~Rosenblatt},
\newblock \bibinfo{title}{The perceptron: a probabilistic model for information
  storage and organization in the brain.},
\newblock \bibinfo{journal}{Psychological review} \bibinfo{volume}{65}
  (\bibinfo{year}{1958}) \bibinfo{pages}{386}.
\bibitem[{Villani(2009)}]{villani2009wasserstein}
\bibinfo{author}{C.~Villani},
\newblock \bibinfo{title}{The wasserstein distances},
\newblock in: \bibinfo{booktitle}{Optimal Transport},
  \bibinfo{publisher}{Springer}, \bibinfo{year}{2009}, pp.
  \bibinfo{pages}{93--111}.
\bibitem[{Laule et~al.(2007)Laule, Vavasour, Kolind, Traboulsee, Moore, Li, and
  MacKay}]{laule2007long}
\bibinfo{author}{C.~Laule}, \bibinfo{author}{I.~M. Vavasour},
  \bibinfo{author}{S.~H. Kolind}, \bibinfo{author}{A.~L. Traboulsee},
  \bibinfo{author}{G.~Moore}, \bibinfo{author}{D.~K. Li},
  \bibinfo{author}{A.~L. MacKay},
\newblock \bibinfo{title}{Long t2 water in multiple sclerosis: What else can we
  learn from multi-echo t2 relaxation?},
\newblock \bibinfo{journal}{Journal of neurology} \bibinfo{volume}{254}
  (\bibinfo{year}{2007}) \bibinfo{pages}{1579--1587}.
\bibitem[{Wansapura et~al.(1999)Wansapura, Holland, Dunn, and
  Ball~Jr}]{wansapura1999nmr}
\bibinfo{author}{J.~P. Wansapura}, \bibinfo{author}{S.~K. Holland},
  \bibinfo{author}{R.~S. Dunn}, \bibinfo{author}{W.~S. Ball~Jr},
\newblock \bibinfo{title}{Nmr relaxation times in the human brain at 3.0
  tesla},
\newblock \bibinfo{journal}{Journal of Magnetic Resonance Imaging: An Official
  Journal of the International Society for Magnetic Resonance in Medicine}
  \bibinfo{volume}{9} (\bibinfo{year}{1999}) \bibinfo{pages}{531--538}.
\bibitem[{Ramdas et~al.(2017)Ramdas, Trillos, and
  Cuturi}]{ramdas2017wasserstein}
\bibinfo{author}{A.~Ramdas}, \bibinfo{author}{N.~G. Trillos},
  \bibinfo{author}{M.~Cuturi},
\newblock \bibinfo{title}{On wasserstein two-sample testing and related
  families of nonparametric tests},
\newblock \bibinfo{journal}{Entropy} \bibinfo{volume}{19}
  (\bibinfo{year}{2017}) \bibinfo{pages}{47}.
\bibitem[{Abadi et~al.(2015)Abadi, Agarwal, Barham, Brevdo, Chen, Citro,
  Corrado, Davis, Dean, Devin, Ghemawat, Goodfellow, Harp, Irving, Isard, Jia,
  Jozefowicz, Kaiser, Kudlur, Levenberg, Man\'{e}, Monga, Moore, Murray, Olah,
  Schuster, Shlens, Steiner, Sutskever, Talwar, Tucker, Vanhoucke, Vasudevan,
  Vi\'{e}gas, Vinyals, Warden, Wattenberg, Wicke, Yu, and
  Zheng}]{tensorflow2015-whitepaper}
\bibinfo{author}{M.~Abadi}, \bibinfo{author}{A.~Agarwal},
  \bibinfo{author}{P.~Barham}, \bibinfo{author}{E.~Brevdo},
  \bibinfo{author}{Z.~Chen}, \bibinfo{author}{C.~Citro}, \bibinfo{author}{G.~S.
  Corrado}, \bibinfo{author}{A.~Davis}, \bibinfo{author}{J.~Dean},
  \bibinfo{author}{M.~Devin}, \bibinfo{author}{S.~Ghemawat},
  \bibinfo{author}{I.~Goodfellow}, \bibinfo{author}{A.~Harp},
  \bibinfo{author}{G.~Irving}, \bibinfo{author}{M.~Isard},
  \bibinfo{author}{Y.~Jia}, \bibinfo{author}{R.~Jozefowicz},
  \bibinfo{author}{L.~Kaiser}, \bibinfo{author}{M.~Kudlur},
  \bibinfo{author}{J.~Levenberg}, \bibinfo{author}{D.~Man\'{e}},
  \bibinfo{author}{R.~Monga}, \bibinfo{author}{S.~Moore},
  \bibinfo{author}{D.~Murray}, \bibinfo{author}{C.~Olah},
  \bibinfo{author}{M.~Schuster}, \bibinfo{author}{J.~Shlens},
  \bibinfo{author}{B.~Steiner}, \bibinfo{author}{I.~Sutskever},
  \bibinfo{author}{K.~Talwar}, \bibinfo{author}{P.~Tucker},
  \bibinfo{author}{V.~Vanhoucke}, \bibinfo{author}{V.~Vasudevan},
  \bibinfo{author}{F.~Vi\'{e}gas}, \bibinfo{author}{O.~Vinyals},
  \bibinfo{author}{P.~Warden}, \bibinfo{author}{M.~Wattenberg},
  \bibinfo{author}{M.~Wicke}, \bibinfo{author}{Y.~Yu},
  \bibinfo{author}{X.~Zheng}, \bibinfo{title}{{TensorFlow}: Large-scale machine
  learning on heterogeneous systems}, \bibinfo{year}{2015}.
  \bibinfo{note}{Software available from tensorflow.org}.
\bibitem[{Van~Rossum et~al.(2000)}]{van2000python}
\bibinfo{author}{G.~Van~Rossum}, et~al., \bibinfo{title}{Python reference
  manual}, \bibinfo{year}{2000}.
\bibitem[{Kingma and Ba(2014)}]{kingma2014adam}
\bibinfo{author}{D.~P. Kingma}, \bibinfo{author}{J.~Ba},
\newblock \bibinfo{title}{Adam: A method for stochastic optimization},
\newblock \bibinfo{journal}{arXiv preprint arXiv:1412.6980}
  (\bibinfo{year}{2014}).
\bibitem[{{Virtanen} et~al.(2020){Virtanen}, {Gommers}, {Oliphant},
  {Haberland}, {Reddy}, {Cournapeau}, {Burovski}, {Peterson}, {Weckesser},
  {Bright}, {van der Walt}, {Brett}, {Wilson}, {Jarrod Millman}, {Mayorov},
  {Nelson}, {Jones}, {Kern}, {Larson}, {Carey}, {Polat}, {Feng}, {Moore}, {Vand
  erPlas}, {Laxalde}, {Perktold}, {Cimrman}, {Henriksen}, {Quintero}, {Harris},
  {Archibald}, {Ribeiro}, {Pedregosa}, {van Mulbregt}, and
  {Contributors}}]{2020SciPy-NMeth}
\bibinfo{author}{P.~{Virtanen}}, \bibinfo{author}{R.~{Gommers}},
  \bibinfo{author}{T.~E. {Oliphant}}, \bibinfo{author}{M.~{Haberland}},
  \bibinfo{author}{T.~{Reddy}}, \bibinfo{author}{D.~{Cournapeau}},
  \bibinfo{author}{E.~{Burovski}}, \bibinfo{author}{P.~{Peterson}},
  \bibinfo{author}{W.~{Weckesser}}, \bibinfo{author}{J.~{Bright}},
  \bibinfo{author}{S.~J. {van der Walt}}, \bibinfo{author}{M.~{Brett}},
  \bibinfo{author}{J.~{Wilson}}, \bibinfo{author}{K.~{Jarrod Millman}},
  \bibinfo{author}{N.~{Mayorov}}, \bibinfo{author}{A.~R.~J. {Nelson}},
  \bibinfo{author}{E.~{Jones}}, \bibinfo{author}{R.~{Kern}},
  \bibinfo{author}{E.~{Larson}}, \bibinfo{author}{C.~{Carey}},
  \bibinfo{author}{{\.I}.~{Polat}}, \bibinfo{author}{Y.~{Feng}},
  \bibinfo{author}{E.~W. {Moore}}, \bibinfo{author}{J.~{Vand erPlas}},
  \bibinfo{author}{D.~{Laxalde}}, \bibinfo{author}{J.~{Perktold}},
  \bibinfo{author}{R.~{Cimrman}}, \bibinfo{author}{I.~{Henriksen}},
  \bibinfo{author}{E.~A. {Quintero}}, \bibinfo{author}{C.~R. {Harris}},
  \bibinfo{author}{A.~M. {Archibald}}, \bibinfo{author}{A.~H. {Ribeiro}},
  \bibinfo{author}{F.~{Pedregosa}}, \bibinfo{author}{P.~{van Mulbregt}},
  \bibinfo{author}{S.~.~. {Contributors}},
\newblock \bibinfo{title}{{SciPy 1.0: Fundamental Algorithms for Scientific
  Computing in Python}},
\newblock \bibinfo{journal}{Nature Methods} \bibinfo{volume}{17}
  (\bibinfo{year}{2020}) \bibinfo{pages}{261--272}.
\bibitem[{Cohen-Adad et~al.(2020)Cohen-Adad, Does, DUVAL, Dyrby, Fieremans,
  Foias, Nami, Sepehrband, Stikov, Zaimi, and et~al.}]{Cohen}
\bibinfo{author}{J.~Cohen-Adad}, \bibinfo{author}{M.~Does},
  \bibinfo{author}{T.~DUVAL}, \bibinfo{author}{T.~B. Dyrby},
  \bibinfo{author}{E.~Fieremans}, \bibinfo{author}{A.~Foias},
  \bibinfo{author}{H.~Nami}, \bibinfo{author}{F.~Sepehrband},
  \bibinfo{author}{N.~Stikov}, \bibinfo{author}{A.~Zaimi},
  \bibinfo{author}{et~al.}, \bibinfo{title}{White matter microscopy database},
  \bibinfo{year}{2020}.
\bibitem[{Vuong et~al.(2017)Vuong, Duval, Cohen-Adad, and
  Stikov}]{bib:Vuong:2017}
\bibinfo{author}{M.-T. Vuong}, \bibinfo{author}{T.~Duval},
  \bibinfo{author}{J.~Cohen-Adad}, \bibinfo{author}{N.~Stikov},
\newblock \bibinfo{title}{On the precision of myelin imaging: Characterizing ex
  vivo dog spinal cord.},
\newblock p. \bibinfo{pages}{3760}.
\bibitem[{Zaimi et~al.(2018)Zaimi, Wabartha, Herman, Antonsanti, Perone, and
  Cohen-Adad}]{zaimi2018axondeepseg}
\bibinfo{author}{A.~Zaimi}, \bibinfo{author}{M.~Wabartha},
  \bibinfo{author}{V.~Herman}, \bibinfo{author}{P.-L. Antonsanti},
  \bibinfo{author}{C.~S. Perone}, \bibinfo{author}{J.~Cohen-Adad},
\newblock \bibinfo{title}{Axondeepseg: automatic axon and myelin segmentation
  from microscopy data using convolutional neural networks},
\newblock \bibinfo{journal}{Scientific reports} \bibinfo{volume}{8}
  (\bibinfo{year}{2018}) \bibinfo{pages}{1--11}.
\bibitem[{Piredda et~al.(2020)Piredda, Hilbert, Canales-Rodrígez, Pizzolato,
  Meuli, Pfeuffer, Thiran, and Kober}]{Piredda}
\bibinfo{author}{G.~F. Piredda}, \bibinfo{author}{T.~Hilbert},
  \bibinfo{author}{E.~J. Canales-Rodrígez}, \bibinfo{author}{M.~Pizzolato},
  \bibinfo{author}{R.~Meuli}, \bibinfo{author}{J.~Pfeuffer},
  \bibinfo{author}{J.-P. Thiran}, \bibinfo{author}{T.~Kober},
\newblock \bibinfo{title}{Fast and high-resolution myelin water imaging:
  Accelerating multi-echo grase with caipirinha},
\newblock \bibinfo{journal}{Magnetic Resonance in Medicine,}
  (\bibinfo{year}{2020}).
\bibitem[{Brant-Zawadzki et~al.(1992)Brant-Zawadzki, Gillan, and
  Nitz}]{brant1992mp}
\bibinfo{author}{M.~Brant-Zawadzki}, \bibinfo{author}{G.~D. Gillan},
  \bibinfo{author}{W.~R. Nitz},
\newblock \bibinfo{title}{Mp rage: a three-dimensional, t1-weighted,
  gradient-echo sequence--initial experience in the brain.},
\newblock \bibinfo{journal}{Radiology} \bibinfo{volume}{182}
  (\bibinfo{year}{1992}) \bibinfo{pages}{769--775}.
\bibitem[{De~Coene et~al.(1992)De~Coene, Hajnal, Gatehouse, Longmore, White,
  Oatridge, Pennock, Young, and Bydder}]{de1992mr}
\bibinfo{author}{B.~De~Coene}, \bibinfo{author}{J.~V. Hajnal},
  \bibinfo{author}{P.~Gatehouse}, \bibinfo{author}{D.~B. Longmore},
  \bibinfo{author}{S.~J. White}, \bibinfo{author}{A.~Oatridge},
  \bibinfo{author}{J.~Pennock}, \bibinfo{author}{I.~Young},
  \bibinfo{author}{G.~Bydder},
\newblock \bibinfo{title}{Mr of the brain using fluid-attenuated inversion
  recovery (flair) pulse sequences.},
\newblock \bibinfo{journal}{American journal of neuroradiology}
  \bibinfo{volume}{13} (\bibinfo{year}{1992}) \bibinfo{pages}{1555--1564}.
\bibitem[{La~Rosa et~al.(2020)La~Rosa, Abdulkadir, Fartaria, Rahmanzadeh, Lu,
  Galbusera, Barakovic, Thiran, Granziera, and Cuadra}]{la2020multiple}
\bibinfo{author}{F.~La~Rosa}, \bibinfo{author}{A.~Abdulkadir},
  \bibinfo{author}{M.~J. Fartaria}, \bibinfo{author}{R.~Rahmanzadeh},
  \bibinfo{author}{P.-J. Lu}, \bibinfo{author}{R.~Galbusera},
  \bibinfo{author}{M.~Barakovic}, \bibinfo{author}{J.-P. Thiran},
  \bibinfo{author}{C.~Granziera}, \bibinfo{author}{M.~B. Cuadra},
\newblock \bibinfo{title}{Multiple sclerosis cortical and wm lesion
  segmentation at 3t mri: a deep learning method based on flair and mp2rage},
\newblock \bibinfo{journal}{NeuroImage: Clinical}  (\bibinfo{year}{2020})
  \bibinfo{pages}{102335}.
\bibitem[{Levesque et~al.(2010)Levesque, Giacomini, Narayanan, Ribeiro, Sled,
  Arnold, and Pike}]{levesque2010quantitative}
\bibinfo{author}{I.~R. Levesque}, \bibinfo{author}{P.~S. Giacomini},
  \bibinfo{author}{S.~Narayanan}, \bibinfo{author}{L.~T. Ribeiro},
  \bibinfo{author}{J.~G. Sled}, \bibinfo{author}{D.~L. Arnold},
  \bibinfo{author}{G.~B. Pike},
\newblock \bibinfo{title}{Quantitative magnetization transfer and myelin water
  imaging of the evolution of acute multiple sclerosis lesions},
\newblock \bibinfo{journal}{Magnetic resonance in medicine}
  \bibinfo{volume}{63} (\bibinfo{year}{2010}) \bibinfo{pages}{633--640}.
\bibitem[{Kolind et~al.(2009)Kolind, M{\"a}dler, Fischer, Li, and
  MacKay}]{kolind2009myelin}
\bibinfo{author}{S.~H. Kolind}, \bibinfo{author}{B.~M{\"a}dler},
  \bibinfo{author}{S.~Fischer}, \bibinfo{author}{D.~K. Li},
  \bibinfo{author}{A.~L. MacKay},
\newblock \bibinfo{title}{Myelin water imaging: implementation and development
  at 3.0 t and comparison to 1.5 t measurements},
\newblock \bibinfo{journal}{Magnetic Resonance in Medicine: An Official Journal
  of the International Society for Magnetic Resonance in Medicine}
  \bibinfo{volume}{62} (\bibinfo{year}{2009}) \bibinfo{pages}{106--115}.
\bibitem[{Laule et~al.(2008)Laule, Kozlowski, Leung, Li, MacKay, and
  Moore}]{laule2008myelin}
\bibinfo{author}{C.~Laule}, \bibinfo{author}{P.~Kozlowski},
  \bibinfo{author}{E.~Leung}, \bibinfo{author}{D.~K. Li},
  \bibinfo{author}{A.~L. MacKay}, \bibinfo{author}{G.~W. Moore},
\newblock \bibinfo{title}{Myelin water imaging of multiple sclerosis at 7 t:
  correlations with histopathology},
\newblock \bibinfo{journal}{Neuroimage} \bibinfo{volume}{40}
  (\bibinfo{year}{2008}) \bibinfo{pages}{1575--1580}.
\bibitem[{Alonso-Ortiz et~al.(2017)Alonso-Ortiz, Levesque, Paquin, and
  Pike}]{alonso2017field}
\bibinfo{author}{E.~Alonso-Ortiz}, \bibinfo{author}{I.~R. Levesque},
  \bibinfo{author}{R.~Paquin}, \bibinfo{author}{G.~B. Pike},
\newblock \bibinfo{title}{Field inhomogeneity correction for gradient echo
  myelin water fraction imaging},
\newblock \bibinfo{journal}{Magnetic resonance in medicine}
  \bibinfo{volume}{78} (\bibinfo{year}{2017}) \bibinfo{pages}{49--57}.
\bibitem[{Oishi et~al.(2008)Oishi, Zilles, Amunts, Faria, Jiang, Li, Akhter,
  Hua, Woods, Toga et~al.}]{oishi2008human}
\bibinfo{author}{K.~Oishi}, \bibinfo{author}{K.~Zilles},
  \bibinfo{author}{K.~Amunts}, \bibinfo{author}{A.~Faria},
  \bibinfo{author}{H.~Jiang}, \bibinfo{author}{X.~Li},
  \bibinfo{author}{K.~Akhter}, \bibinfo{author}{K.~Hua},
  \bibinfo{author}{R.~Woods}, \bibinfo{author}{A.~W. Toga}, et~al.,
\newblock \bibinfo{title}{Human brain white matter atlas: identification and
  assignment of common anatomical structures in superficial white matter},
\newblock \bibinfo{journal}{Neuroimage} \bibinfo{volume}{43}
  (\bibinfo{year}{2008}) \bibinfo{pages}{447--457}.
\bibitem[{Mori et~al.(2008)Mori, Oishi, Jiang, Jiang, Li, Akhter, Hua, Faria,
  Mahmood, Woods et~al.}]{mori2008stereotaxic}
\bibinfo{author}{S.~Mori}, \bibinfo{author}{K.~Oishi},
  \bibinfo{author}{H.~Jiang}, \bibinfo{author}{L.~Jiang},
  \bibinfo{author}{X.~Li}, \bibinfo{author}{K.~Akhter},
  \bibinfo{author}{K.~Hua}, \bibinfo{author}{A.~V. Faria},
  \bibinfo{author}{A.~Mahmood}, \bibinfo{author}{R.~Woods}, et~al.,
\newblock \bibinfo{title}{Stereotaxic white matter atlas based on diffusion
  tensor imaging in an icbm template},
\newblock \bibinfo{journal}{Neuroimage} \bibinfo{volume}{40}
  (\bibinfo{year}{2008}) \bibinfo{pages}{570--582}.
\bibitem[{Adler and {\"O}ktem(2017)}]{adler2017solving}
\bibinfo{author}{J.~Adler}, \bibinfo{author}{O.~{\"O}ktem},
\newblock \bibinfo{title}{Solving ill-posed inverse problems using iterative
  deep neural networks},
\newblock \bibinfo{journal}{Inverse Problems} \bibinfo{volume}{33}
  (\bibinfo{year}{2017}) \bibinfo{pages}{124007}.
\bibitem[{Aja-Fern{\'a}ndez and
  Vegas-S{\'a}nchez-Ferrero(2016)}]{aja2016statistical}
\bibinfo{author}{S.~Aja-Fern{\'a}ndez},
  \bibinfo{author}{G.~Vegas-S{\'a}nchez-Ferrero},
\newblock \bibinfo{title}{Statistical analysis of noise in mri},
\newblock \bibinfo{journal}{Switzerland: Springer International Publishing}
  (\bibinfo{year}{2016}).
\bibitem[{Sled(2018)}]{sled2018modelling}
\bibinfo{author}{J.~G. Sled},
\newblock \bibinfo{title}{Modelling and interpretation of magnetization
  transfer imaging in the brain},
\newblock \bibinfo{journal}{Neuroimage} \bibinfo{volume}{182}
  (\bibinfo{year}{2018}) \bibinfo{pages}{128--135}.
\bibitem[{Malik et~al.(2018)Malik, Teixeira, and Hajnal}]{malik2018extended}
\bibinfo{author}{S.~J. Malik}, \bibinfo{author}{R.~P.~A. Teixeira},
  \bibinfo{author}{J.~V. Hajnal},
\newblock \bibinfo{title}{Extended phase graph formalism for systems with
  magnetization transfer and exchange},
\newblock \bibinfo{journal}{Magnetic resonance in medicine}
  \bibinfo{volume}{80} (\bibinfo{year}{2018}) \bibinfo{pages}{767--779}.
\bibitem[{Wang et~al.(2020)Wang, Ostenson, and Smith}]{wang2020snapmrf}
\bibinfo{author}{D.~Wang}, \bibinfo{author}{J.~Ostenson},
  \bibinfo{author}{D.~S. Smith},
\newblock \bibinfo{title}{snapmrf: Gpu-accelerated magnetic resonance
  fingerprinting dictionary generation and matching using extended phase
  graphs},
\newblock \bibinfo{journal}{Magnetic Resonance Imaging} \bibinfo{volume}{66}
  (\bibinfo{year}{2020}) \bibinfo{pages}{248--256}.
\bibitem[{Does et~al.(2019)Does, Olesen, Harkins, Serradas-Duarte, Gochberg,
  Jespersen, and Shemesh}]{does2019evaluation}
\bibinfo{author}{M.~D. Does}, \bibinfo{author}{J.~L. Olesen},
  \bibinfo{author}{K.~D. Harkins}, \bibinfo{author}{T.~Serradas-Duarte},
  \bibinfo{author}{D.~F. Gochberg}, \bibinfo{author}{S.~N. Jespersen},
  \bibinfo{author}{N.~Shemesh},
\newblock \bibinfo{title}{Evaluation of principal component analysis image
  denoising on multi-exponential mri relaxometry},
\newblock \bibinfo{journal}{Magnetic resonance in medicine}
  \bibinfo{volume}{81} (\bibinfo{year}{2019}) \bibinfo{pages}{3503--3514}.
\bibitem[{Bouhrara et~al.(2018)Bouhrara, Reiter, Maring, Bonny, and
  Spencer}]{bouhrara2018use}
\bibinfo{author}{M.~Bouhrara}, \bibinfo{author}{D.~A. Reiter},
  \bibinfo{author}{M.~C. Maring}, \bibinfo{author}{J.-M. Bonny},
  \bibinfo{author}{R.~G. Spencer},
\newblock \bibinfo{title}{Use of the nesma filter to improve myelin water
  fraction mapping with brain mri},
\newblock \bibinfo{journal}{Journal of Neuroimaging} \bibinfo{volume}{28}
  (\bibinfo{year}{2018}) \bibinfo{pages}{640--649}.
\bibitem[{El-Hajj et~al.(2020)El-Hajj, Moussaoui, Collewet, and
  Musse}]{el2020multi}
\bibinfo{author}{C.~El-Hajj}, \bibinfo{author}{S.~Moussaoui},
  \bibinfo{author}{G.~Collewet}, \bibinfo{author}{M.~Musse},
\newblock \bibinfo{title}{Multi-exponential transverse relaxation times
  estimation from magnetic resonance images under rician noise and spatial
  regularization},
\newblock \bibinfo{journal}{IEEE Transactions on Image Processing}
  (\bibinfo{year}{2020}).
\bibitem[{Hwang and Du(2009)}]{hwang2009improved}
\bibinfo{author}{D.~Hwang}, \bibinfo{author}{Y.~P. Du},
\newblock \bibinfo{title}{Improved myelin water quantification using spatially
  regularized non-negative least squares algorithm},
\newblock \bibinfo{journal}{Journal of Magnetic Resonance Imaging: An Official
  Journal of the International Society for Magnetic Resonance in Medicine}
  \bibinfo{volume}{30} (\bibinfo{year}{2009}) \bibinfo{pages}{203--208}.
\bibitem[{Kumar et~al.(2018)Kumar, Hariharan, Faizy, Borchert, Siemonsen,
  Fiehler, Reddy, and Sedlacik}]{kumar2018using}
\bibinfo{author}{D.~Kumar}, \bibinfo{author}{H.~Hariharan},
  \bibinfo{author}{T.~D. Faizy}, \bibinfo{author}{P.~Borchert},
  \bibinfo{author}{S.~Siemonsen}, \bibinfo{author}{J.~Fiehler},
  \bibinfo{author}{R.~Reddy}, \bibinfo{author}{J.~Sedlacik},
\newblock \bibinfo{title}{Using 3d spatial correlations to improve the noise
  robustness of multi component analysis of 3d multi echo quantitative t2
  relaxometry data},
\newblock \bibinfo{journal}{NeuroImage} \bibinfo{volume}{178}
  (\bibinfo{year}{2018}) \bibinfo{pages}{583--601}.
\bibitem[{Nagtegaal et~al.(2020)Nagtegaal, Koken, Amthor, de~Bresser,
  M{\"a}dler, Vos, and Doneva}]{nagtegaal2020myelin}
\bibinfo{author}{M.~Nagtegaal}, \bibinfo{author}{P.~Koken},
  \bibinfo{author}{T.~Amthor}, \bibinfo{author}{J.~de~Bresser},
  \bibinfo{author}{B.~M{\"a}dler}, \bibinfo{author}{F.~Vos},
  \bibinfo{author}{M.~Doneva},
\newblock \bibinfo{title}{Myelin water imaging from multi-echo t2 mr
  relaxometry data using a joint sparsity constraint},
\newblock \bibinfo{journal}{NeuroImage}  (\bibinfo{year}{2020})
  \bibinfo{pages}{117014}.
\bibitem[{Nguyen et~al.(2012)Nguyen, Wisnieff, Cooper, Kumar, Raj,
  Spincemaille, Wang, Vartanian, and Gauthier}]{nguyen2012t2prep}
\bibinfo{author}{T.~D. Nguyen}, \bibinfo{author}{C.~Wisnieff},
  \bibinfo{author}{M.~A. Cooper}, \bibinfo{author}{D.~Kumar},
  \bibinfo{author}{A.~Raj}, \bibinfo{author}{P.~Spincemaille},
  \bibinfo{author}{Y.~Wang}, \bibinfo{author}{T.~Vartanian},
  \bibinfo{author}{S.~A. Gauthier},
\newblock \bibinfo{title}{T2prep three-dimensional spiral imaging with
  efficient whole brain coverage for myelin water quantification at 1.5 tesla},
\newblock \bibinfo{journal}{Magnetic Resonance in Medicine}
  \bibinfo{volume}{67} (\bibinfo{year}{2012}) \bibinfo{pages}{614--621}.
\bibitem[{Tang et~al.(2018)Tang, Fernandez-Granda, Lannuzel, Bernstein,
  Lattanzi, Cloos, Knoll, and Assl{\"a}nder}]{tang2018multicompartment}
\bibinfo{author}{S.~Tang}, \bibinfo{author}{C.~Fernandez-Granda},
  \bibinfo{author}{S.~Lannuzel}, \bibinfo{author}{B.~Bernstein},
  \bibinfo{author}{R.~Lattanzi}, \bibinfo{author}{M.~Cloos},
  \bibinfo{author}{F.~Knoll}, \bibinfo{author}{J.~Assl{\"a}nder},
\newblock \bibinfo{title}{Multicompartment magnetic resonance fingerprinting},
\newblock \bibinfo{journal}{Inverse problems} \bibinfo{volume}{34}
  (\bibinfo{year}{2018}) \bibinfo{pages}{094005}.
\bibitem[{McGivney et~al.(2018)McGivney, Deshmane, Jiang, Ma, Badve, Sloan,
  Gulani, and Griswold}]{mcgivney2018bayesian}
\bibinfo{author}{D.~McGivney}, \bibinfo{author}{A.~Deshmane},
  \bibinfo{author}{Y.~Jiang}, \bibinfo{author}{D.~Ma},
  \bibinfo{author}{C.~Badve}, \bibinfo{author}{A.~Sloan},
  \bibinfo{author}{V.~Gulani}, \bibinfo{author}{M.~Griswold},
\newblock \bibinfo{title}{Bayesian estimation of multicomponent relaxation
  parameters in magnetic resonance fingerprinting},
\newblock \bibinfo{journal}{Magnetic resonance in medicine}
  \bibinfo{volume}{80} (\bibinfo{year}{2018}) \bibinfo{pages}{159--170}.
\bibitem[{Hoppe et~al.(2017)Hoppe, K{\"o}rzd{\"o}rfer, W{\"u}rfl, Wetzl,
  Lugauer, Pfeuffer, and Maier}]{hoppe2017deep}
\bibinfo{author}{E.~Hoppe}, \bibinfo{author}{G.~K{\"o}rzd{\"o}rfer},
  \bibinfo{author}{T.~W{\"u}rfl}, \bibinfo{author}{J.~Wetzl},
  \bibinfo{author}{F.~Lugauer}, \bibinfo{author}{J.~Pfeuffer},
  \bibinfo{author}{A.~K. Maier},
\newblock \bibinfo{title}{Deep learning for magnetic resonance fingerprinting:
  A new approach for predicting quantitative parameter values from time
  series.},
\newblock in: \bibinfo{booktitle}{GMDS}, pp. \bibinfo{pages}{202--206}.
\bibitem[{Hochreiter and Schmidhuber(1997)}]{hochreiter1997long}
\bibinfo{author}{S.~Hochreiter}, \bibinfo{author}{J.~Schmidhuber},
\newblock \bibinfo{title}{Long short-term memory},
\newblock \bibinfo{journal}{Neural computation} \bibinfo{volume}{9}
  (\bibinfo{year}{1997}) \bibinfo{pages}{1735--1780}.

\end{thebibliography}

\end{document}